\PassOptionsToPackage{table,xcdraw}{xcolor}

\documentclass[letterpaper,twocolumn,10pt]{article}
\usepackage{usenix2019_v3}

\usepackage{hyperref}
\usepackage{multirow}
\usepackage{booktabs}
\usepackage{amsmath}
\usepackage{textcomp}
\usepackage{color, colortbl}
\usepackage{graphicx,subfigure}
\usepackage{cleveref}
\usepackage{pifont}
\usepackage{gensymb}
\usepackage[ruled]{algorithm2e} 
\usepackage{enumitem}
\usepackage{authblk}
\usepackage{makecell}
\usepackage{threeparttable}

\newcommand{\name}{Full-Duplex LoRa Backscatter }
\newcommand{\shortname}{FD LoRa Backscatter }
\newcommand{\xref}[1]{\S\ref{#1}}
\newcommand{\squishlist}{\begin{itemize}[itemsep=1pt,parsep=2pt,topsep=3pt,partopsep=0pt,leftmargin=0em, itemindent=1em,labelwidth=1em,labelsep=0.5em]}
\newcommand{\squishend}{\end{itemize}}
\newcommand{\squishenum}{\begin{enumerate}[itemsep=1pt,parsep=2pt,topsep=3pt,partopsep=0pt,leftmargin=0em,listparindent=1.5em,labelwidth=1em,labelsep=0.5em]}
\newcommand{\squishsubenum}{\begin{enumerate}[itemsep=1pt,parsep=2pt,topsep=0pt,partopsep=0pt,leftmargin=0em,listparindent=1.5em,labelwidth=1em,labelsep=0.5em]}
\newcommand{\squishenumend}{\end{enumerate}}

\newcommand\numberthis{\addtocounter{equation}{1}\tag{\theequation}}

\newcommand{\capsize}{\footnotesize}
\newcommand{\rev}{}
\newcommand{\revf}{}
\newcommand{\cmt}[1]{\ignorespaces}

\usepackage{lipsum} 
\usepackage{fancyhdr}
\begin{document}
\setlength{\belowdisplayskip}{0pt} 
\setlength{\belowdisplayshortskip}{0pt} 
\setlength{\abovedisplayskip}{0pt} 
\setlength{\abovedisplayshortskip}{0pt}
\date{}
\pagenumbering{gobble}

\title{\Large \bf Simplifying Backscatter Deployment: Full-Duplex LoRa Backscatter}


\author[1,2\thanks{Work was done while the author was at Jeeva Wireless.}]{\rm Mohamad Katanbaf}
\author[1]{\rm Anthony Weinand}
\author[1]{\rm Vamsi Talla}
\affil[1]{Jeeva Wireless}
\affil[2]{University of Washington}

\maketitle

\begin{abstract}

Due to the practical challenges in the deployment of existing half-duplex systems, the promise of ubiquitous backscatter connectivity has eluded us. To address this, we design the first long-range full-duplex LoRa backscatter reader. We leverage existing LoRa chipsets as transceivers and use a microcontroller in combination with inexpensive passive elements including a hybrid coupler, inductors, tunable capacitors, and resistors to achieve 78~dB of self-interference cancellation and build a low-cost, long-range, and small-form-factor \name reader. 

We evaluate our system in various deployments and show that we can successfully communicate with a backscatter tag at distances of up to 300~ft in line of sight, and through obstacles, such as walls and cubicles, in a 4,000~ft$^2$ office area. 
We reconfigure our reader to conform to the size and power requirements of a smartphone, and demonstrate communication with a contact-lens-form-factor prototype device.
Finally, we attach our reader to a drone and demonstrate backscatter sensing for precision agriculture with an instantaneous coverage of 7,850~ft$^2$.

\end{abstract}
\section{Introduction}
\label{sec:intro}
{\rev 
Recent advances~\cite{AmbientBackscatter, passiveWiFi, InterTechnology, hitchhike, lorabackscatter, plora} have demonstrated the potential of backscatter to replace power-hungry, large, expensive radios~\footnote{active RFIDs are also radios} with an orders of magnitude lower power, smaller-size, cheaper, potentially battery-free connectivity solution. 
This promise, however, has run into practical limitations in regard to existing backscatter infrastructure. 
Full-duplex (FD) RFID readers~\cite{st25ru3993,Speedway} and other proprietary full-duplex systems~\cite{Chen2019, backfi2015}, in which a single reader communicates with tags are easy to deploy, but these existing readers are large, complex, expensive, and have limited range. 

To address this, recent half-duplex (HD) backscatter systems have leveraged the economies of scales and ubiquity of industry-standard protocols such as WiFi~\cite{passiveWiFi,hitchhike}, Bluetooth~\cite{Ensworth2017, freerider, relacks}, ZigBee~\cite{passive-zigbee, freerider}, and LoRa~\cite{lorabackscatter, plora} to reduce the cost, size, and complexity of reader infrastructure and achieve longer range. 
However, these systems suffer from deployment issues, as the half-duplex topology requires two physically-separated radio devices: one for transmitting the carrier, and another for receiving the backscattered data packet. 
The need to deploy multiple devices in different locations significantly limits the use cases for backscatter.

\begin{figure}[!t]
    \centering
    \begin{minipage}{1\linewidth}
        \centering
		\includegraphics[width=1\linewidth]
        {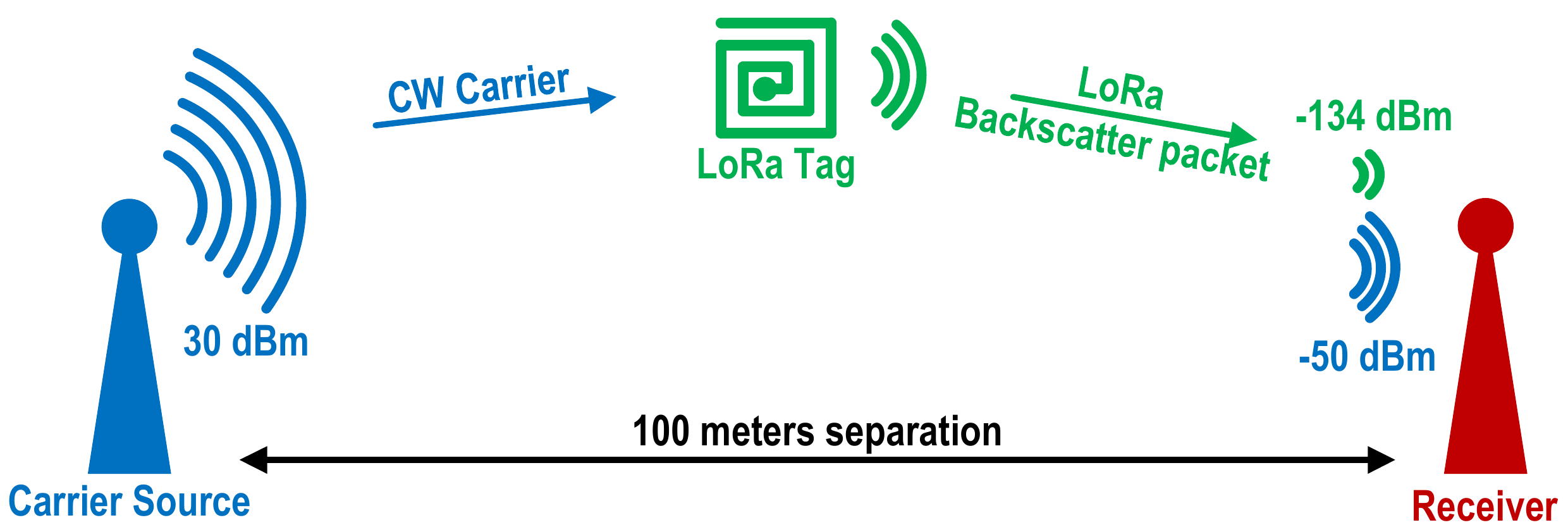}\\
         \vskip -0.05 in
         
        \capsize{(a)  \textbf{Half-Duplex deployment.} The carrier source and receiver are separated by 100m to mitigate carrier interference.}
        \vskip -0.15 in
        \hfill
    \end{minipage}
    \vskip 0.0 in
    \begin{minipage}{1\linewidth}
        \centering
        \includegraphics[width=1\linewidth]
        {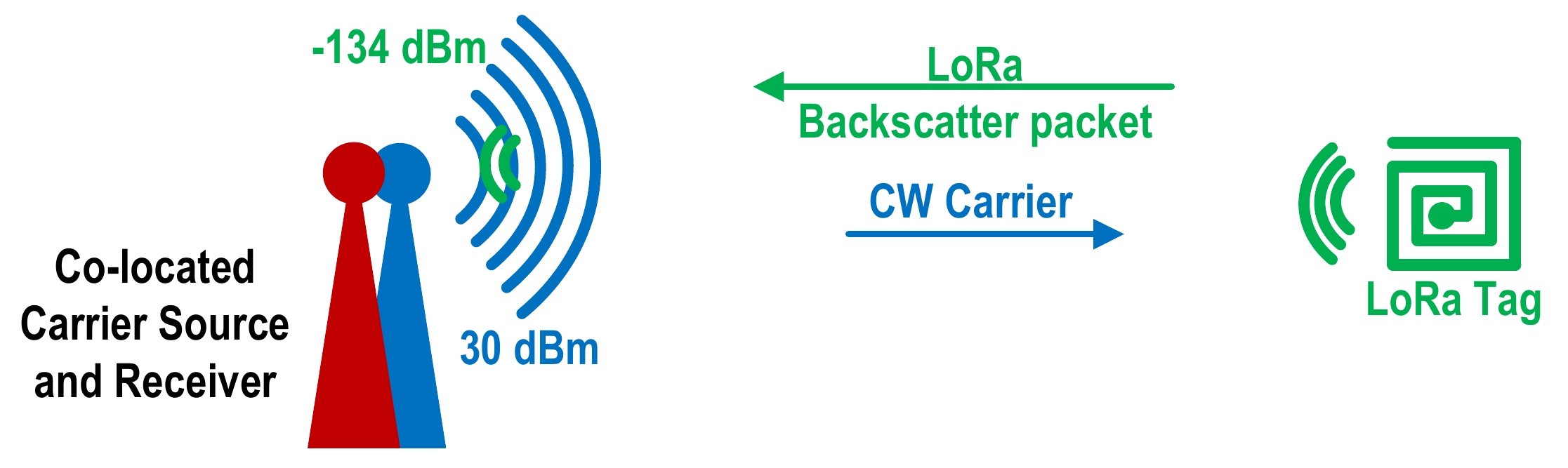}\\
         \vskip -0.05 in
        \capsize{(b) \textbf{Full-Duplex deployment.} The carrier source and receiver are co-located and need 78~dB interference cancellation.}
        \vskip -0.15 in
    \end{minipage}
    \vskip -0.1 in
    \caption{Overview of HD and FD Backscatter Deployments.}
    \label{fig:deployment-intro}
    \vskip -0.15 in
\end{figure}

We present the first \name reader which combines the low-cost, long-range, and small-form-factor benefits of a standard-protocols-compliant backscatter system with the simple deployment of a full-duplex system. 
This addresses one of the remaining pain points of backscatter and opens backscatter to a plethora of applications. 
A low-cost, long-range, small-form-factor full-duplex reader could be easily integrated into existing devices. This would enable peripheral, wearable, and medical devices such as pill bottles~\cite{pill_bottle,Mauro2019}, insulin pens~\cite{insulin_pen,Klonoff2018}, smart glasses~\cite{Klinker2020,Rauschnabel2016}, and contact lenses~\cite{Jeon2020,mojovision} to use backscatter to directly communicate with a \cmt {single reader integrated into a} \cmt{ portable device such as a} cellphone, tablet, or laptop. Similarly, in agriculture, aerial surveillance drones could be equipped with a backscatter reader to collect data from sensors distributed throughout a field.} 

{\rev To understand the challenge in building an \shortname reader, consider the traditional HD LoRa backscatter system, shown in Fig.~\ref{fig:deployment-intro}(a).} The first radio transmits a single-tone carrier at power levels up to 30~dBm. 
A tag uses subcarrier modulation to backscatter a packet at an offset frequency, which is then received by the second radio, 100~m from the transmitter.
The physical separation is necessary to attenuate the out-of-band carrier at the receiver to a level where it does not impact the sensitivity~\cite{lorabackscatter}. 
This illustrates the fundamental challenge in a \shortname system, as shown in Fig.~\ref{fig:deployment-intro}(b): The single-tone carrier needs {\rev at least 78~dB of suppression}\cmt{attenuation}, a 63-million~$\times$ reduction in signal strength, between the transmitter and a commodity LoRa receiver chipset, both integrated on the same PCB. 
This {\rev suppression} must be implemented in the analog RF domain without substantially increasing the cost, complexity, or power consumption of the system.
Furthermore, unlike path-loss attenuation, which is wide-band, typical cancellation techniques have a trade-off between cancellation depth and bandwidth~\cite{bharadia_full_2013,korpi_compact_2017,liempd_70_2016}.
If the cancellation bandwidth is insufficient, the carrier phase noise will show up as in-band noise at the receiver.
{\rev Therefore}, a second requirement is to bring the phase noise of the carrier at the offset frequency to below the {\rev receiver's} input noise level.

{\rev 
Existing FD cancellation techniques, including analog, digital, and hybrid designs used in full-duplex radios have different or more relaxed requirements and, as a consequence, are insufficient to meet the needs of our system. Analog and hybrid cancellation techniques require bulky and expensive RF components such as circulators~\cite{SKYFR-001400}, vector modulators~\cite{hmc630}, and phase shifters~\cite{JSPHS-1000} to achieve sufficient cancellation, each of which increases cost and size. \cmt {Furthermore, the active components generate additional noise, degrading receiver performance~\cite{chu_integrated_2018}.}
Digital cancellation techniques require access to IQ samples, which are unavailable on commodity radios, and instead use SDRs~\cite{bharadia_full_2013,sim_nonlinear_2017,Chen2019,Akeela2018}, FPGAs~\cite{Tsoeunyane2017}, or DSPs~\cite{Atomix}, which are all prohibitively expensive. For a detailed analysis of why existing techniques are insufficient, see~\xref{sec:related}.}


\cmt{We observe that the scaling of CMOS technology has resulted in an exponential reduction in the size and cost of digital computation.} 
\cmt{ Our key idea is to leverage CMOS scaling and use a cheap ARM microcontroller to control a tunable impedance network to achieve deep cancellation of the single-tone carrier in the analog domain. This allows us to use a commodity radio to build the \shortname system and bring low-cost, long-range, easy-to-deploy backscatter connectivity to a range of new applications.} 

{\rev Our key idea is to leverage the ubiquity and economies of scale of existing LoRa transceivers and microcontrollers and add inexpensive passive components to fulfill the two requirements of full-duplex operation. This enables us to build a \textit{low-cost, long-range, small-form-factor}, \name reader}. We use a single-antenna topology with a hybrid coupler to interface the transmitter and the receiver with the antenna. The leakage from the transmitter to the receiver, i.e. self-interference, is a function of the impedance at the coupled port. A microcontroller adaptively tunes an impedance network, tracking variations in the antenna impedance and environmental reflections with the objective of minimizing interference at the receiver. 

{\rev We introduce a \textit{novel, two-stage, tunable impedance network} to achieve 78~dB suppression of carrier signal.
The extent to which a carrier is suppressed is a function of how closely the tunable impedance can \cmt{is able to} track the variations in antenna impedance}, which in turn depends on the resolution of the impedance network.
{\rev A single-stage network is limited by the step size of its digital capacitors and does not have a high enough resolution to reliably achieve 78~dB cancellation~\cite{Jung2011,Keehr2018,liempd_70_2016, Chu2018}. 
Our two-stage network is built by cascading two stages, each consisting of four, 5-bit tunable capacitors and two fixed inductors with an attenuating resistor network between the stages.}
\cmt{The first stage has enough resolution to reliably achieve at least 50~dB of cancellation. The resistor network has the effect of shrinking the perceived impedance variation of the second stage. We design the stages such that the impedance variation of the second stage, as seen at the coupled port, covers the step size of the first stage, thereby providing fine-grain control of the impedance with no dead zones.} 
{\rev The two-stage design provides the fine-grain control and coverage necessary to meet the 78~dB carrier cancellation target across the expected range of variation in antenna impedance.}
To achieve the second objective of bringing the phase noise of the carrier at the offset frequency below the noise floor of the receiver, while simultaneously obtaining 78~dB cancellation at the carrier frequency, is very challenging. 
There is a fundamental trade-off between the cancellation depth and bandwidth~\cite{bharadia_full_2013,korpi_compact_2017,liempd_70_2016}, and we prioritize the 78~dB cancellation requirement at the carrier frequency. 
We use a COTS synthesizer with low phase noise to relax the cancellation requirement at the offset frequency.


We implement the \name system using only COTS components, for a total cost of \$27.54 at low volumes, only 10\% more than the cost of two HD units. Our evaluation shows that the two-stage impedance network \cmt{was able to} achieve{\rev s} >78~dB carrier cancellation and >46~dB of offset cancellation in practical scenarios with a tuning time overhead of less than $2.7\%$. Results are summarized below:

\squishlist

\item The FD reader can communicate with tags at distances of up to 300~ft in line of sight. When placed in the corner of a 4,000~ft$^2$ office space with concrete, glass, and wood structures and walls, tags can operate within the entire space.

\item \cmt{We build a mobile version for integration into portable devices.}
{\revf We integrate a low-power configuration of the FD reader into portable devices.}
We attach the prototype to the back of a smartphone and show that the tags can communicate at distances beyond 50~ft at 20~dBm, 25~ft at 10~dBm, and up to 20~ft at 4~dBm.

\item We build two proof-of-concept applications\cmt{with the mobile reader}. We prototype a contact-lens-form-factor antenna tag and show that it can communicate with FD reader attached to a smartphone at distances of up to 22~ft and when the reader is inside a user's pocket. We also attach the FD reader to a quadcopter and {\rev fly it to 60~ft above a field}. The reader is able to communicate with tags placed on the ground at a lateral distance of up to 50~ft, corresponding to an instantaneous coverage of 7,850~ft$^2$.
\squishend

\cmt{
\vskip 0.05in\noindent{\bf Case for \name.} In Table~\ref{table:comp_bs}, we provide a summary of the state of the art in backscatter communication and compare it with our \shortname system. Our system achieves 6 × longer range than the best reported FD system and outperforms most HD systems. Only the HD LoRa backscatter~\cite{lorabackscatter} system has a
longer symmetric range at 237.5~m, but that evaluation was based on a -143~dBm sensitivity protocol with 45 bps and 2.4 sec packet length. In our design, we limit receive sensitivity to -134~dBm to ensure that packets are shorter than 400 ms, the maximum channel dwell time for FCC compliant frequency hop-ping systems~\cite{FCC}. Additionally, we use a hybrid-coupler architecture to reduce cost, which incurs a loss of 7 dB (see ~\xref{sec:approach}). So, in total, our link budget is reduced by 16 dB which translates to about 2.5 × range reduction, close to the 90~m range of our system. We believe that this is a reasonable trade-off {\rev for the size and price point of the \shortname system.}}
\section{Background}
\label{sec:background}
{\rev
Our work brings full-duplex operation to a LoRa Backscatter system. We start with a background on LoRa backscatter, followed by a primer on the full-duplex operation.}

\vspace{-2mm}
\subsection{LoRa Backscatter Primer}
\label{sec:system_lora_primer}

{\rev Backscatter communication eliminates RF carrier generation at the tag and, instead, uses switches to reflect existing, ambient RF signals for data transmission.} 
This drastically reduces the cost, size, and power consumption of wireless communication~\cite{passiveWiFi,lorabackscatter, backfi2015}. 
In HD backscatter deployment, as shown in Fig.~\ref{fig:deployment-intro}(a), a radio source generates the single-tone carrier, a backscatter tag reflects the carrier to synthesize LoRa packets, then a receiver decodes the backscattered packets.
{\rev However, the carrier signal also ends up as a strong source of interference at the receiver, which degrades the  receiver's ability to decode packets. Backscatter systems use two key techniques to mitigate carrier interference~\cite{Ensworth2017,freerider,passiveWiFi,InterTechnology,witag,spatialstream,passive-zigbee,lorabackscatter,plora,Netscatter,OFDMAbackscatter,25MbpsBrain,batteryfreevideo,WiFiCam}. 
First, the tag uses subcarrier modulation to synthesize packets at a frequency offset from the carrier frequency. 
The receiver operates at the offset frequency, pushing the interference, i.e. the carrier signal, out of band at the receiver. 
Since receivers are designed to operate in the presence of out-of-band interference, the receiver can decode the backscattered packets with minimal loss in sensitivity. 
Second, the transmitter and receiver are physically separated to attenuate that carrier to a level where it does not affect the receiver's sensitivity\cmt{of the receiver}.
However, in full-duplex systems, by definition, the transmitter and receiver cannot be physically separated.}

LoRa receivers have low sensitivity and high blocker tolerance, making them ideal candidates for long-range backscatter connectivity, as demonstrated by prior HD backscatter designs~\cite{lorabackscatter, plora}.
LoRa has two key protocol parameters that can be used to trade off data rate with receive sensitivity: spreading factor and bandwidth.
Since our system transmits up to 30~dBm, the FCC mandates frequency hopping{\rev and a maximum channel dwell time of 400~ms~\cite{FCC}}.
So, we limit our system to protocols with packet~lengths shorter than this limit, which corresponds to a sensitivity of -134dBm.
Longer packet lengths are incompatible with frequency hopping unless we implement intra-packet frequency hopping. 
Doing so would require tuning in the middle of packet reception, which is not feasible on commodity radio receivers.

\vspace{-2mm}
\subsection{Full-Duplex Primer}
\label{sec:br_fd_primer}
{\rev
FD radios transmit and receive at the same time on a single device, allowing simultaneous communication between devices without delay. The main obstacle in achieving FD functionality is self-interference; the strong transmit signal leaks to the sensitive receiver and appears as interference, degrading its performance. Hence, the key is to suppress the interference before it reaches the receiver. Broadly speaking, there are two approaches to FD operation: out-of-band and in-band.

Cellular standards such as WCDMA and LTE implement out-of-band full-duplexing by using Frequency Division Duplexing (FDD), where two distinct fixed frequency bands are used for uplink and downlink.
In FDD systems, the operating frequency and frequency offset are fixed, and a frequency selective duplexer is used to suppress the transmitter leakage in the receive band.
At first glance, it may look like FD backscatter is similar, as the tag backscatters the data packet at a frequency offset, however, FDD systems use much higher offset frequencies, at least 40~MHz for WCDMA and LTE bands above 800~MHz~\cite{ltebands}, in line with the requirements of practical frequency duplexers and passive filters.
Backscatter systems, in contrast, transmit the carrier and receive the packet within the same frequency band, keeping the frequency offset low to minimize the power consumption of the tag. 
For example, the LoRa backscatter system operating in the 902-928MHz ISM band uses an offset frequency of 2-4~MHz. With such a small offset, we cannot leverage passive filters or frequency duplexers used in FDD systems.

In recent years, researchers have demonstrated success with In-Band Full-Duplex (IBFD) radios \cite{bharadia_full_2013,korpi_compact_2017,duarte_design_2014,Chen2019}, where radios transmit and receive simultaneously on the same frequency.
IBFD radios suppress self-interference by using a combination of analog and digital cancellation techniques to bring the signal strength of the typically-wideband carrier below the noise floor of the receiver over the entire receive bandwidth. 
IBFD radios use isolation and analog cancellation techniques in the RF domain to first bring the carrier signal below the saturation level of the receiver front end.
Next, digital cancellation techniques are used in the baseband to suppress the signal below the noise floor across the receiver bandwidth. 
Since the frequency offset in a backscatter system is small, an FD backscatter system can leverage SI suppression techniques similar to IBFD devices, such as isolation and analog cancellation. 
However, there are two key differences. 
First, the \shortname system uses a single-tone signal as the carrier, so we need to suppress a very narrow-band signal. 
Second, unlike IBFD radios, the \shortname system uses existing COTS radios, which do not provide access to signals in the digital baseband of the receiver. 
Hence, unlike IBFD systems, which use digital cancellation in addition to analog cancellation, we need to achieve 78~dB of SI cancellation entirely within the analog domain.


}

\vspace{-2mm}
\section{\shortname Requirements}
\label{sec:system}






\cmt{Our goal is to combine the carrier source and the receiver into a single low-cost, low-complexity, small-form-factor device. Since we cannot rely on physical separation to attenuate the carrier before it reaches the receiver, we need to incorporate a cancellation network to suppress the carrier and reduce SI at the receiver. } 
In this work, we focus on the cancellation requirements for a LoRa backscatter system, but the design techniques and architecture are not LoRa specific. 
They can be extended to build FD backscatter systems for other wireless standards such as WiFi, Zigbee, Bluetooth, SigFox, or NB-IoT {\rev that use a single-tone carrier and subcarrier modulation to synthesize backscatter packets~\cite{passiveWiFi, Ensworth2017, InterTechnology}. 
However, these techniques are not directly applicable to systems which do not use sub-carrier modulation~\cite{backfi2015} or use wide-band Wi-Fi or LoRa packets as carrier~\cite{freerider,plora,passive-zigbee}.} 

We divide the cancellation requirements into two categories: carrier cancellation and offset cancellation.

\vspace{-2mm}
\subsection{Carrier Cancellation}

\begin{figure}[t!]
\centering
\includegraphics[width=1\columnwidth]{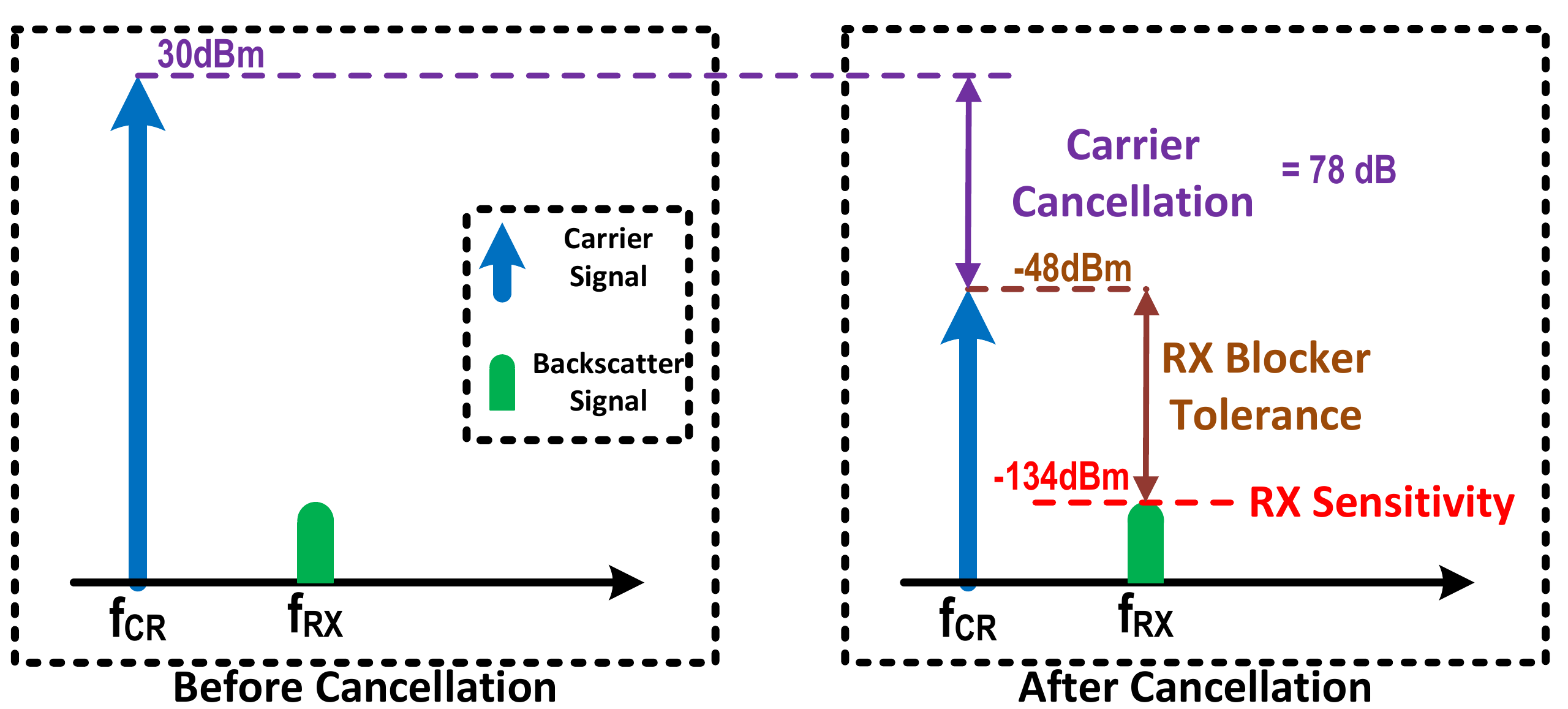}
	\vskip -0.15in
\caption{\textbf{Carrier Cancellation.} 
    \textnormal{
     The 30~dBm single-tone carrier needs 78~dB of attenuation to meet {\rev the Rx} blocker tolerance and ensure packet reception down to Rx sensitivity. }}
\label{fig:car_can_req}
	\vskip -0.2in
\end{figure} 

{\rev
We define carrier cancellation ($CAN_{CR}$) as the required cancellation of the carrier signal (the source of self-interference) at the center frequency. The carrier acts as a \textit{blocker} i.e. a strong signal in the vicinity of the desired signal that can affect a receiver's performance and reduce its sensitivity. A strong blocker can saturate the LNA, forcing it to reduce gain and increase the noise floor. Secondly, post LNA, a blocker can mix with the receiver local-oscillator phase-noise and contribute to in-band noise. Finally, baseband filters have limited stopband attenuation, and even a small portion of the blocker passing through the filter reduces the signal-to-noise and -interference ratio.

We compute the minimum required carrier cancellation as
\begin{align}
CAN_{CR} > P_{CR} - Rx_{Sen} - Rx_{BT} \numberthis 
\label{eq:blocker}
\end{align}

\noindent {where $P_{CR}$ is the carrier power, $Rx_{Sen}$ is the receiver sensitivity, and $Rx_{BT}$ is the receiver blocker tolerance.}

For example, as per the SX1276 datasheet, the blocker tolerance at a 2~MHz offset for $BW=125~kHz$, $SF=12$, -137~dBm sensitivity protocol is 94~dB~\cite{sx1276}. 
Based on equation \ref{eq:blocker}, for $P_{CR} = 30$~dBm, at least 73~dB of SI cancellation is required. 
However, the datasheet blocker specification assumes a 3~dB degradation in sensitivity, which is substantial for backscatter systems.
Additionally, the datasheet provides blocker specifications for only a subset of frequency offsets and protocol parameters. 
To get a more comprehensive set of requirements, we perform our own blocker experiments for a range of frequency offsets (2, 3, and 4~MHz) and protocol parameters (366~bps~-~13.6~kbps data rates). 
We connect a LoRa transmitter and a single-tone generator to two variable attenuators and combine their outputs at a LoRa receiver.
First, we get a baseline sensitivity by turning off the single tone and increase the LoRa transmitter attenuation till we reach receiver sensitivity, defined by PER~<~10\%. 
Next, we turn on the single-tone generator with maximum attenuation and reduce the attenuation, thereby increasing blocker power, until the PER exceeds 10\%. 
We record the maximum tolerable interference power for different frequency offsets, receiver bandwidths, and spreading factors to achieve different data rates and conclude that 78~dB is the most stringent carrier-cancellation specification. 
As mentioned prior, the blocker performance of a receiver depends on both the RF front end and digital baseband.
Our analysis shows that the SX1276 baseband has sufficient digital baseband filtering to suppress the blocker at the offset frequency, and additional filtering would not help in this specific case.}

Fig.~\ref{fig:car_can_req} illustrates the carrier cancellation requirement. 
Before cancellation, the carrier signal is much stronger, but after cancellation, the SI is dropped to a level that the receiver can tolerate. The difference between these levels is $CAN_{CR}$.
Note that a lower cancellation may suffice for some data rates and frequency offsets, but our design supports all configurations.

\vspace{-2mm}
\subsection{Offset Cancellation} 

\begin{figure}[t!]
\centering
\includegraphics[width=1\columnwidth]{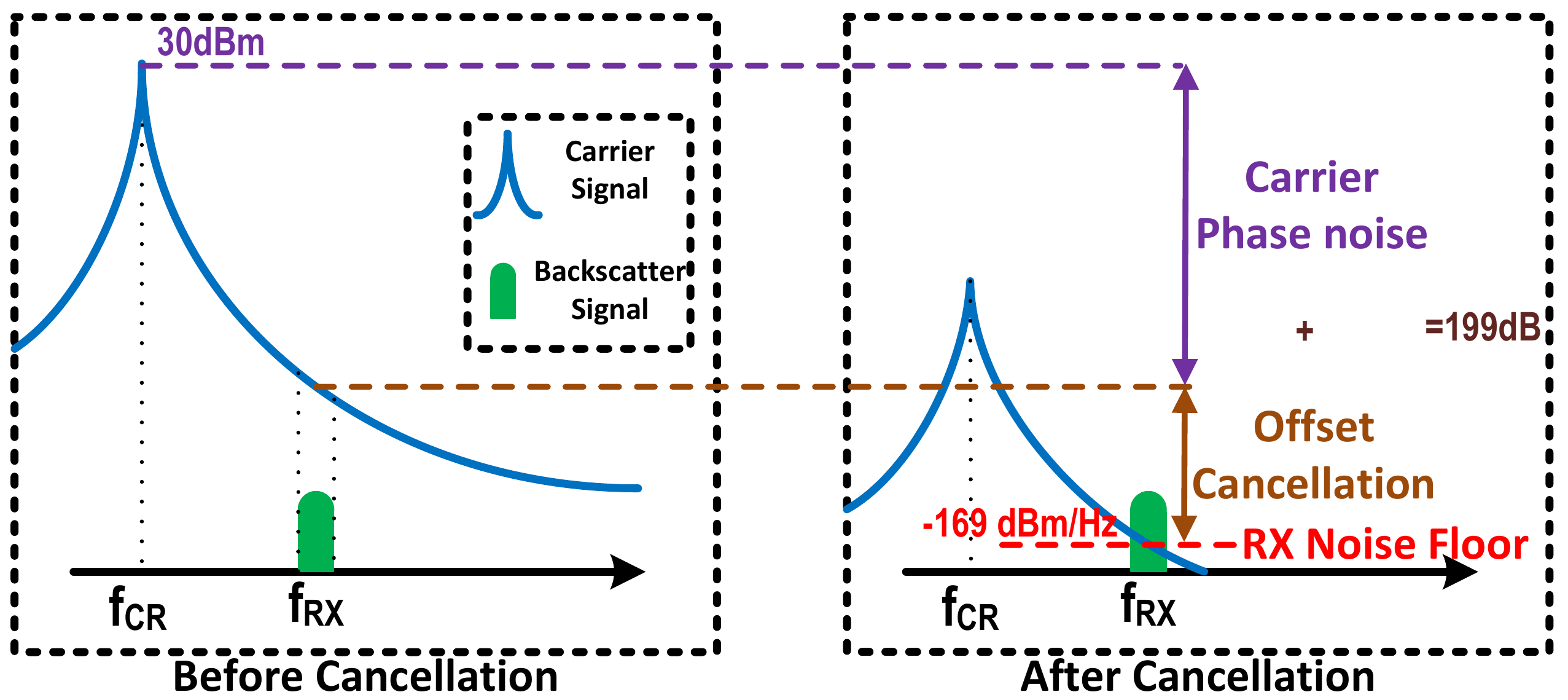}
	\vskip -0.15in
   \caption{\textbf{Offset Cancellation}. 
    \textnormal{ The phase noise of the single-tone carrier at the frequency offset after cancellation should be less than the noise floor of the receiver.}}
    \label{fig:off_can_req}
	\vskip -0.2in
\end{figure} 

{\rev
We define offset cancellation ($CAN_{OFS}$) as cancellation of the carrier signal at the offset frequency. We use a single-tone signal as the carrier. An ideal oscillator generates a pure sine wave, {\rev with the entire} signal power concentrated at a single frequency. However, practical oscillators have phase-modulated noise components, which spreads the power to adjacent frequencies. This results in noise side bands~\cite{razavi_rf_2012} and is characterized by the source's phase noise, i.e. power spectral density (dBc/Hz) of the noise at a frequency offset from the center frequency. Since the receiver operates at a frequency offset from the carrier, the carrier phase noise shows up as in-band noise to the receiver. 
For optimal receiver performance, the SI signal after cancellation at the offset frequency should be less than the receiver noise floor.
We compute the minimum required offset cancellation as 

\newcommand{\Lagr}{\mathcal{L}}
\begin{align*}
&P_{CR} + \Lagr_{CR(\Delta f)} + 10Log(B) - CAN_{OFS} < 10Log(KTB) + Rx_{NF} \\ \nonumber
&CAN_{OFS} - \Lagr_{CR(\Delta f)} > P_{CR} - 10Log(KT) - Rx_{NF} \numberthis \label{phn}
\end{align*}

\noindent{where $\Lagr_{CR(\Delta f)}$ is the carrier phase noise at the offset frequency ($\Delta f$), $B$ is the receiver bandwidth, $K$ is the Boltzmann constant, $T$ is temperature, and $Rx_{NF}$ is receiver noise figure.} We show this requirement in Fig.~\ref{fig:off_can_req}. Before cancellation, the backscattered signal is buried under the carrier phase noise, but, after cancellation, the carrier phase noise is pushed below the receiver noise floor. 

As per the SX1276 datasheet $Rx_{NF} = 4.5$~dB, so for $P_{CR} = 30$~dBm, $CAN_{OFS} - \Lagr_{CR(\Delta f)} > 199.5$~dB.
{\rev The offset cancellation depends on transmit power, carrier phase noise, and receiver noise figure. Interestingly, since the channel bandwidth appears on both sides of equation~\ref{phn}, offset cancellation is independent of the receiver channel bandwidth. In other words, we can use the carrier phase noise per unit bandwidth and receiver noise floor per unit bandwidth instead of the \cmt{accumulated} aggregate noise over the entire bandwidth.}

In an HD system, the transmitter and receiver are physically separated, and the carrier attenuation via RF propagation does not significantly change with frequency. 
So, if the 78~dB carrier cancellation requirement is met, the phase noise at the offset frequency would also be attenuated by the same amount. 
However, cancellation networks do not have the same wide-band frequency characteristics~\cite{Dinc2015,Reiskarimian2018,Keehr2018,liempd_70_2016, Chu2018}. 
The inequality shows that {\rev satisfying the offset cancellation requirement for an FD system} requires a joint design of the carrier source and the cancellation network. If we use a high-phase-noise carrier, we would need high offset cancellation, whereas if we lower the phase noise of the carrier source, we can relax the offset cancellation requirements. 
}
Carrier and offset SI cancellation are functions of offset frequency, and both typically relax with an increase in offset frequency. However, an increase in offset frequency increases the tag power consumption. Thus, the frequency offset presents a trade-off between tag power consumption and SI cancellation requirements. 2-4~MHz is a reasonable compromise; we use a 3~MHz offset frequency in this work.

{\revf \section{\shortname Reader Design}}
\label{sec:approach}

\begin{figure}[!t]
    \centering
    \includegraphics[width=1\columnwidth]{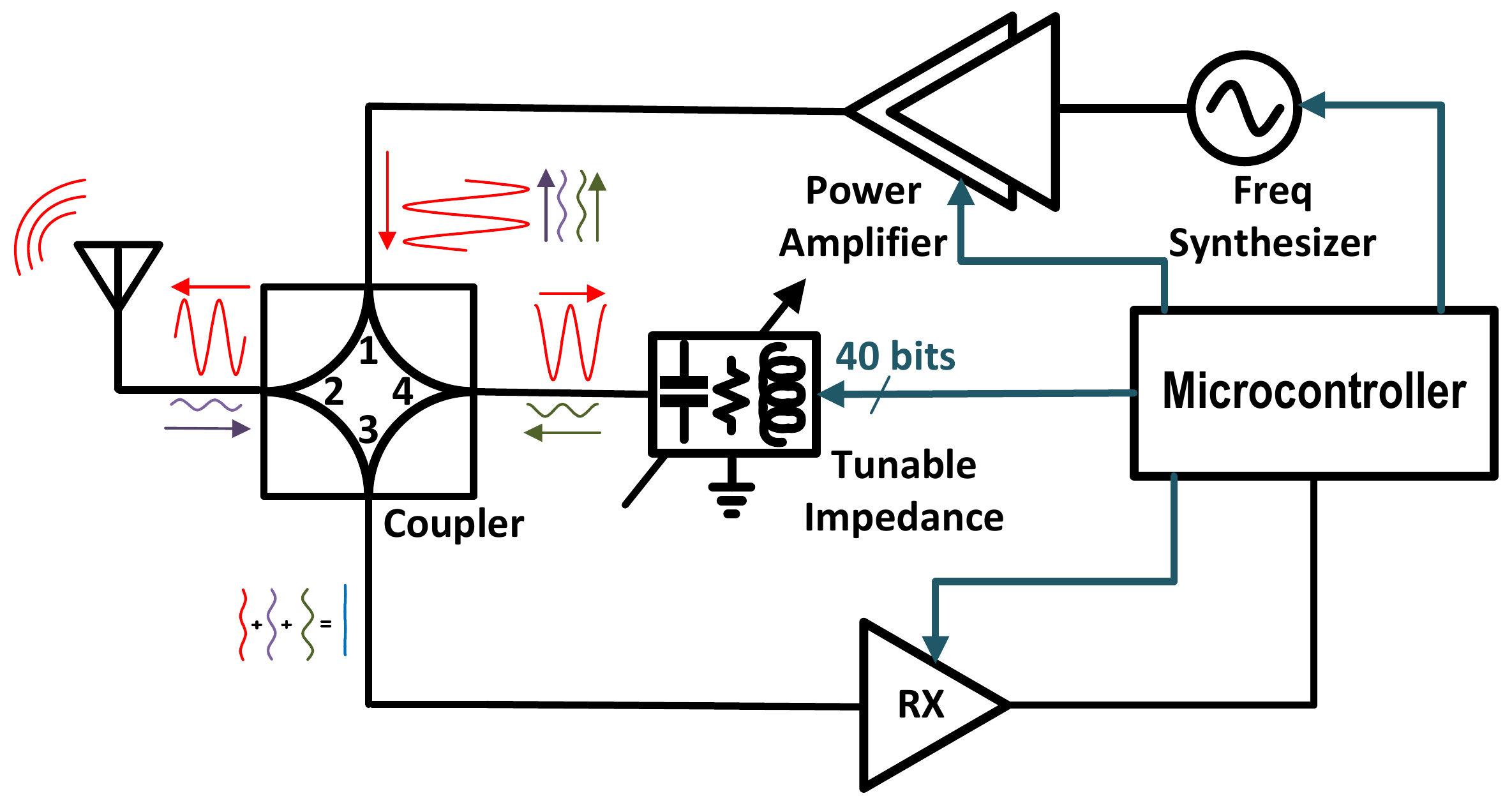}
	\vskip -0.15in
    \caption{\textbf{System Architecture.}  
    \textnormal{
         We use a single-antenna hybrid-coupler architecture with a two-stage tunable impedance network for cancellation. A microcontroller controls all components and implements the tuning algorithm.}}
    \label{fig:block_diagram}
    \vskip -0.2 in
\end{figure}

The \shortname system uses a single antenna and a hybrid coupler with a two-stage tunable impedance network to achieve carrier and offset cancellation. The cancellation network uses only passive components, minimizing cost, complexity, power consumption, and noise. Fig.~\ref{fig:block_diagram} shows the block diagram of our design. The antenna is connected to the transmit and receive paths via a hybrid coupler. The fourth port of the coupler is connected to a tunable impedance network that tracks the antenna impedance to reflect and phase shift a portion of the single-tone carrier, suppressing the SI that flows to the receiver. The carrier signal is generated by a frequency synthesizer and fed to a power amplifier (PA) to transmit up to 30~dBm. An on-board microcontroller controls the synthesizer, PA, receiver radio, and digital tunable impedance network using a SPI interface. We use RSSI readings from the receiver to measure the SI there.

Below, we describe the principle of operation for the hybrid coupler. This is followed by the two-stage tunable impedance network design and the choice of carrier source to meet the carrier and offset cancellation requirements. Finally, we describe the tuning algorithm.

\begin{figure*}[t!]
    \centering
    \vskip -0.1 in
    \begin{minipage}{.49\linewidth}
        \centering
        \begin{minipage}{1\linewidth}
        \centering
	        \capsize
	        \includegraphics[width=1\linewidth]
	        {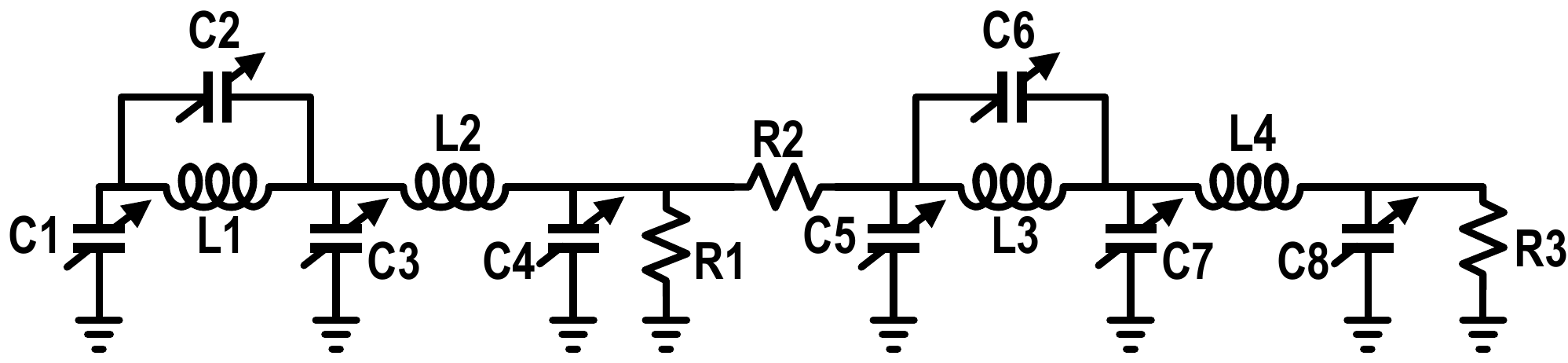}\\
	        	\vskip -0.05 in
	         (a) Topology of the two stage impedance network.
	        \vskip -0.1 in
	        \hfill
        \end{minipage}
        \begin{minipage}{1\linewidth}
        \centering
	        \capsize
	        \includegraphics[width=1\linewidth]
	        {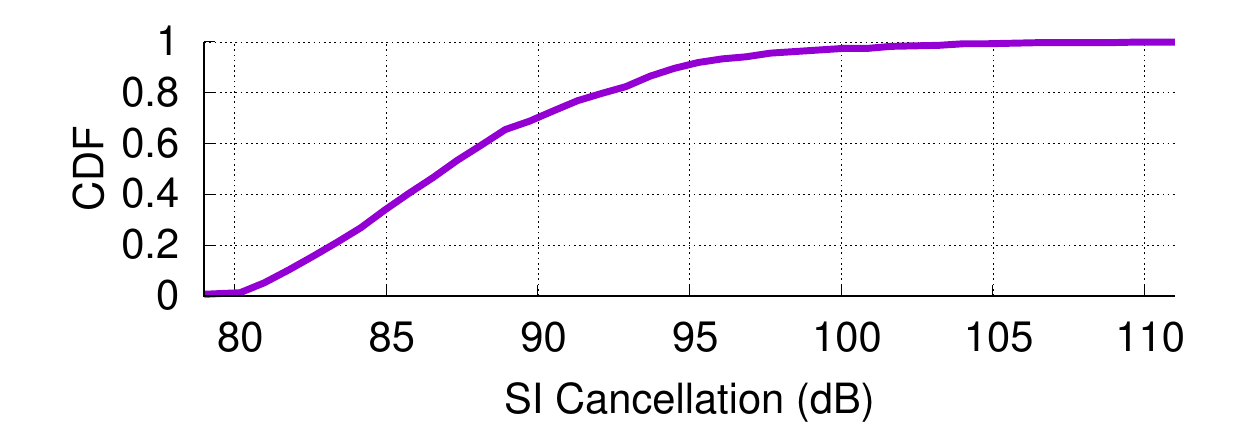}\\
	        	        	\vskip -0.075 in
	         (b) Simulated SI cancellation for 400 random antenna impedance 
	        \vskip -0.1 in
	        \hfill
        \end{minipage}
    \end{minipage}
    \begin{minipage}{.245\linewidth}
        \centering
        \capsize
        \includegraphics[width=1\linewidth]{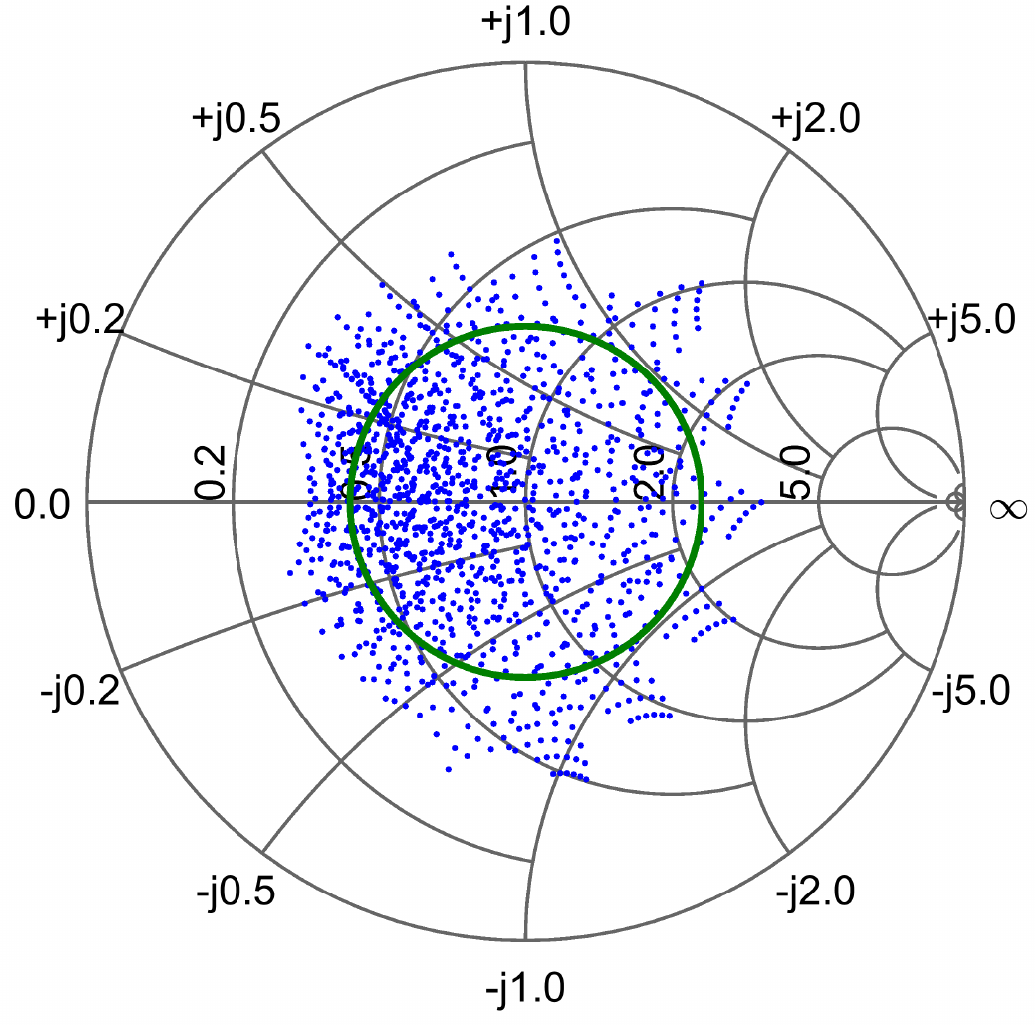}\\
        	        	\vskip -0.05 in
        (c) \textbf{Tunable impedance coverage.} 
        We show the coverage of the first stage on smith chart. 
        
        (step size = 6~LSBs) 
        \vskip -0.11 in
        \hfill
    \end{minipage}
    \begin{minipage}{0.225\linewidth}
        \centering
        \capsize
        \includegraphics[width=1\linewidth]{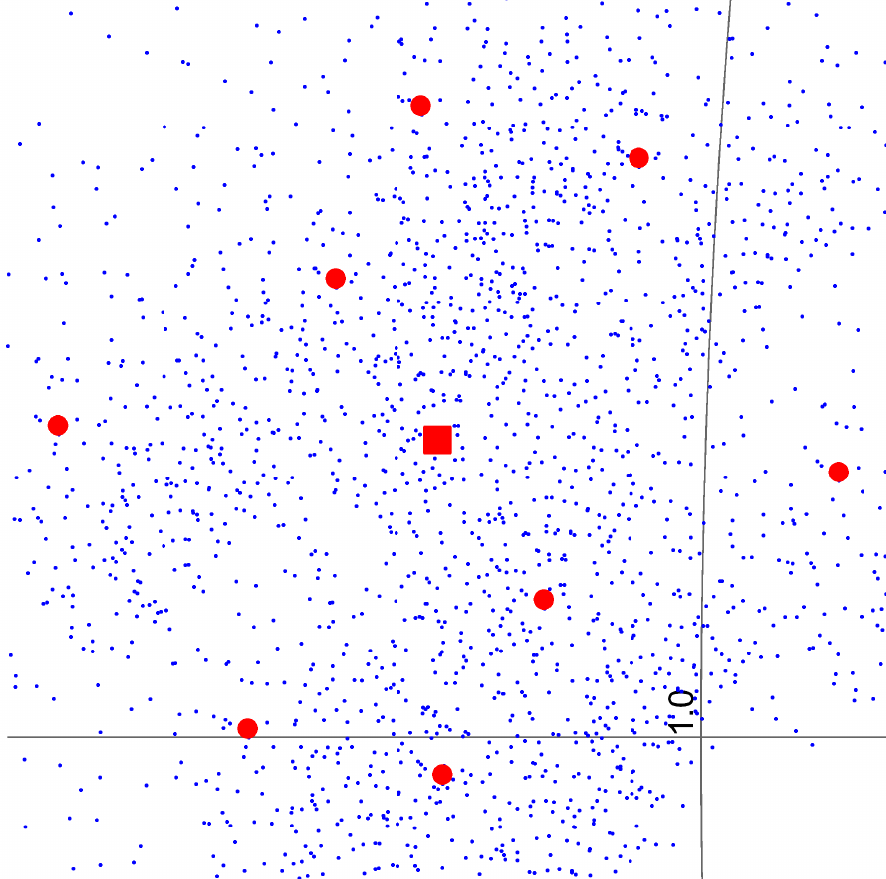}\\
        \vskip -0.05 in
        (d) \textbf{Second stage fine tuning.} We show impedance states with finer resolution achieved using second stage(blue dots)
        
        (step size = 10~LSBs)
        \vskip -0.1 in
    \end{minipage}
    \vskip -0.1 in
    \caption{Two stage tunable impedance network used for SI cancellation and its simulation results. }
    \label{fig:tunable_imp}
    \vskip -0.25 in
\end{figure*}

\vspace{-2mm}
\subsection{Coupler: Principle of Operation}\label{coupler}
Couplers are four-port devices that divide an incident signal between the output and coupled port while keeping the fourth port isolated~\cite{Parad1965, Pon1961}; we use this property to isolate the transmitter and receiver. 
We connect the transmitter to the input port~(1), the receiver to the isolated port~(3), the antenna to the output port~(2), and the tunable impedance network to the coupled port~(4). 
The carrier signal is split between the antenna and coupled ports, leaving the receiver isolated. 
The received signal at the antenna port is split between the receiver and the transmitter, leaving the tunable impedance isolated.
Couplers are reciprocal elements, meaning that the signal split described above is symmetrical. Ideally, we would want the entire PA output to go to the antenna (low TX insertion loss) and the entire received signal from the antenna to go to the receiver (low RX insertion loss). A higher coupling factor would direct more of the PA output to the antenna at the cost of reducing signal transmission from the antenna to the receiver. Since our goal is to maximize the communication range, we must minimize the sum of {\rev the} transmit and receive insertion losses. A hybrid, or 3~dB coupler, equally divides the input power between the output and coupled ports and minimizes total loss to 6~dB.

Two factors limit the practical SI cancellation of a hybrid coupler. First, every coupler has leakage: a typical COTS coupler provides $\sim$~25~dB of isolation between the transmit and receive ports~\cite{X3C09P1}, far lower than our requirement. Second, and more important, is the effect of the antenna. Typical antennas, including low-cost PIFAs, are characterized by -10~dB return loss, and any reflection from the antenna port would couple to the receiver and further contribute to the SI.


\vskip 0.05in\noindent{\bf Antenna Impedance Variation.} Environmental variations affect antenna impedance, i.e. nearby objects can detune the antenna or create additional reflections that contribute to variation in its reflection coefficient. 
Since SI cancellation is dependent on antenna impedance, it is {\rev essential} \cmt{important} to characterize the expected variation. We design a 1.9~in~$\times$~0.8~in PIFA for our implementation and use this antenna to quantify impedance variation. 
We connect the PIFA to an Agilent 8753ES VNA~\cite{our_vna} and subject it to environmental variations by placing the antenna upright on a table, and measure $S_{11}$ as a hand and other objects approach it from different directions. Our results show that the magnitude of {\rev the} reflection coefficient, $|\Gamma|$, reaches a maximum of 0.38, and we set expected $|\Gamma|~<~0.4$ for the antenna.

\vspace{-2mm}
\subsection{Two-Stage Tunable Impedance Network}
{\rev
We use a tunable impedance at the coupled port~\cite{Jung2011,Keehr2018,liempd_70_2016} to negate SI leakage due to variation in antenna impedance.
However, the cancellation depth is a function of how precisely we can tune this impedance. To achieve 78~dB carrier cancellation, we need a very fine resolution for the tunable impedance, which is not available in COTS digital capacitors. To solve this, we introduce a novel two-stage tunable impedance network that allows us to achieve deep cancellation, using only passive components.

To understand how {\rev the tunable impedance network improves the SI suppression from the coupler}, follow the flow of signals in Fig.~\ref{fig:block_diagram}. The carrier (red) splits between the antenna port and the coupled port. The impedance at the coupled port is tuned such that the reflection from this port (green) cancels out both the leakage of the coupler and the reflection from the antenna port (purple) to achieve deep cancellation at the receive port (blue). In the worst case of a significantly detuned antenna, reflection from the antenna is much stronger than the leakage, and this should be canceled by the reflection from the tunable impedance. 

We use a topology of two fixed inductors and four digitally tunable capacitors terminated with a resistor to cover the range of expected impedance values required to cancel strong reflections from the antenna~\cite{liempd_70_2016}. 
We observe that in a tuning network terminated with a resistor, only a small portion of the signal is reflected, and most of it is dissipated. We replace the termination resistor with a resistive signal divider, followed by a second tunable impedance, as shown in Fig.~\ref{fig:tunable_imp}(a). The signal reflected by the second stage flows through the resistive divider twice, effectively lowering the impact of impedance changes in the second stage on the overall reflected signal. 
This allows us to use the second stage to make much more fine changes in impedance, increasing the tuning resolution and enabling deep cancellation. 
However, {\rev the first stage still determines} the tuning range and offset cancellation. \cmt{are still determined by the first stage.}
The second stage provides the fine resolution to accurately match the reflection from the balanced port with the leakage and reflection from the antenna port.
To eliminate dead zones, we choose the resistive divider such that the fine tuning network covers the step size of the coarse tuning network.
This coarse and fine tuning approach is similar to using a hybrid coupler combined with an analog feed-forward path. 
The second stage effectively provides the functionality of a feed-forward path, but without additional noise and at a lower cost, lower complexity, and lower power. 

}
%

We simulate the behavior of our tunable impedance network to demonstrate the efficacy of our approach. We plot the tuning network reflection coefficient at 915~MHz in Fig.~\ref{fig:tunable_imp}~(c) for a step size of six LSBs for each capacitor in the first stage. For visibility, the plot only shows 1,296 impedance states out of more than 1 million first-stage impedance states and more than 1 trillion total states. The plot demonstrates that our design can cover the impedances corresponding to the antenna reflection coefficient circle of $|\Gamma|~<~0.4$. Next, we show the fine tuning of the second stage in Fig.~\ref{fig:tunable_imp}~(d). We select an initial state (the large, red square in the center) and change each capacitor in the first stage by one LSB to get the other eight red dots. Without the second stage, we would not be able to achieve any impedance between these dots. For each of these nine states, we use a step size of 10 LSBs for each capacitor in the second stage and show the resulting impedances in blue. The blue cloud shows the fine resolution control covering the dead zone between the first-stage steps. Finally, we simulate SI cancellation for 400 random points of antenna impedance inside the $|\Gamma|~<~0.4$ circle and plot the cancellation CDF in Fig.~\ref{fig:tunable_imp}~(b). Cancellation of $>80$~dB is achieved for the 1st percentile, which satisfies our requirement.

\vspace{-2mm}
\subsection{Carrier Source Selection} {\rev
The phase noise of the carrier source determines the required cancellation at the offset frequency, as shown in equation~\ref{phn}. In order to understand the requirements of the carrier source, we first simulate SI cancellation at different frequency offsets. We tune the capacitors to achieve a minimum of 78~dB carrier cancellation and then sweep the operating frequency. 
Our results show that at 3~MHz frequency offset, 47~dB cancellation or more is achieved in 90\% of the simulated cases.
If we use SX1276 with a phase noise of -130~dBc at 3~MHz offset as our transmitter, then 47~dB of offset cancellation is insufficient. 

The tuning network has multiple poles that can be optimized to increase cancellation bandwidth \cite{liempd_70_2016,Khaledian2018}. However, doing so reduces the cancellation at the center frequency. This approach would also require a wide-band receiver to provide feedback on the SI power over the entire bandwidth to tune the capacitors, which is unavailable on the SX1276 (max. BW = 500~kHz). We need $>78$~dB of SI cancellation at the carrier frequency and prioritize this requirement. Instead of SX1276, we use ADF4351~\cite{ADF4351} frequency synthesizer to generate the single tone carrier, which has a lower phase noise of -153~dBc at 3~MHz offset. Although the ADF4351 is slightly more expensive, it relaxes the offset cancellation requirement to 46.5~dB and eliminates the need for an additional wide-band receiver or a power detector, justifying the design choice.
}

\vspace{-2mm}
\subsection{Tuning Algorithm}
\label{tuning}

Our design uses a two-stage impedance network with eight digital capacitors, each with five control bits; a total of 40 bits resulting in $2^{40}$ ($\sim$~1-trillion) states for the impedance network. Multiple capacitor states can result in the impedance required for 78~dB cancellation, and any one of those is acceptable. As it is impossible to search across such a vast space in real-time, there is a need for an efficient tuning algorithm that can run on a commodity ARM Cortex-M4 microcontroller. 

We use a simulated annealing algorithm to tune the capacitors in each stage separately~\cite{annealing}. Simulated annealing is based on the physical process of heating, and then slowly cooling, a material to minimize defects in its structure. For every temperature value, we take ten steps. At each step, we add a random value bounded by a maximum step size to each capacitor and measure the SI using receiver RSSI measurement. We accept the new capacitor values if the SI decreases, or with a temperature-dependent probability when the SI increases. We start with a temperature equal to 512 and divide it by two each round until it reaches one. We set predefined cancellation thresholds for each stage and stop the tuning once the thresholds are met. If the first stage reaches the threshold (set to 50~dB), but the second stage fails to do so, we repeat the tuning until either it converges or reaches a timeout.

\section{Implementation}
\label{sec:impl}


We implement the \shortname reader for operation in 902-928~MHz on a 3.8~in~$\times$~1.9~in, 4-layer, FR4 PCB. 
We {\rev place} \cmt{incorporate} the RF components, including antenna, transmitter, receiver, and cancellation network, on the top side of the PCB, and microcontroller and power management on the bottom.
We use the SX1276 as the LoRa receiver~\cite{sx1276}. 
The cancellation circuit consists of the X3C09P1 $90\degree$ hybrid coupler~\cite{X3C09P1} and a two-stage tunable impedance network,{\rev shown in Fig.~\ref{fig:tunable_imp}~(a).} 
Variable capacitors $C_{1}$-$C_{8}$ are implemented by pSemi PE64906 tunable capacitors, with 32 linear steps from 0.9~pF~-~4.6~pF~\cite{PE64906}. 
We set inductors $L_{1}$, $L_{3}$ to 3.9~nH and $L_{2}$, $L_{4}$ to 3.6~nH.
We set resistors $R_{1}$, $R_{2}$, and $R_{3}$ to $62~\Omega$, $240~\Omega$, and $50~\Omega$ respectively. 
We use the ADF4351 synthesizer to generate the single-tone carrier, which has 23~dB better phase noise at 3~MHz offset compared to the SX1276. 
The output power of the carrier can be amplified up to 30~dBm using the SKY65313-21 power amplifier~\cite{sky65313}. 
Our cancellation technique has an expected loss of 7-8~dB; 6~dB of which is the theoretical loss due to hybrid coupler architecture, the rest is due to component non-idealities.

We design a custom coplanar inverted-F PCB antenna. \cmt{at minimal additional cost.} The radiating element of the antenna measures 1.9~in~$\times$~0.8~in and leverages the existing ground plane for omnidirectional radiation. We measure the performance of the antenna in an anechoic chamber, and results show a peak gain of 1.2~dB, half-power beam-width of 80\degree, and cumulative efficiency of 78\%. The transmitter, receiver, and cancellation network are controlled using a SPI interface by an on-board ARM Cortex-M4 STM32F4 microcontroller~\cite{stm32f413rg}. The microcontroller implements a state machine in C to transition between tuning, downlink, and uplink operating modes. In the tuning mode, the microcontroller first configures the center frequency and power of the carrier and then tunes the impedance network to minimize SI using the simulated annealing algorithm described in~\xref{tuning}. After the tuning phase, the MCU sends the downlink OOK message to wake up the backscatter tag. Then, it transitions to the uplink mode where it configures the receiver with the appropriate LoRa protocol parameters to decode backscattered packets. The MCU then repeats this cycle for the next frequency.

\vspace{-2mm}
{\revf \subsection{FD Reader Configurations}}
\label{sec:power}

\noindent {\revf We configure the \shortname reader for two different use cases; a `base-station' setup and a `mobile' setup. Below we describe each configuration.}

\vskip 0.05in\noindent{\bf Base-Station Configuration:} The base-station configuration of the \shortname reader uses a 8~dBc high gain patch antenna~\cite{patchantenna}. The synthesizer and PA are set to transmit at 30~dBm. These settings maximize operating range and we use this configuration for the line-of-sight and non-line-of-sight range testing. In the base-station setup method, the power amplifier, synthesizer, receiver, and MCU consume 2,580~mW, 380~mW, 40~mW, and 40~mW, respectively, resulting in total power consumption of 3.04~W.
3.04~W is not a limitation for a plugged-in device such as a smart speaker or WiFi router, but is too high for a portable device.

\vskip 0.05in
\noindent{\bf Mobile Configuration:}
For applications with lower power consumption and smaller size requirements, we configured the system as a `mobile' version. 
We use the on-board antenna and configure the synthesizer and PA to transmit at lower power levels of 4~dBm, 10~dBm, and 20~dBm.
Since the PA and synthesizer dominate power consumption, reducing transmit power greatly reduces power consumption.
In this mobile configuration, power consumption is low enough to draw from conventional portable power sources like a USB battery or a laptop.
It is also small enough that, if desired, we are able to attach it to an iPhone 11 Pro without increasing the phone's footprint, shown in Fig.~\ref{fig:Phone}(a).

{\revf Lower transmit powers relax cancellation requirements (see ~\xref{sec:approach}), which can be leveraged to further reduce the power consumption of the synthesizer and the power amplifier. For 20~dBm output power, we can instead use an LMX2571~\cite{lmx2571} as the synthesizer which has higher phase noise, but lower power consumption. 
We can also use a CC1910~\cite{cc1190} as the PA which operates efficiently at 20~dBm output power. 
Similarly, for output powers of 4~dBm and 10~dBm, we can use a CC1310~\cite{cc1310} as the synthesizer and eliminate the PA. 
These optimizations would reduce power consumption to levels shown in Table~\ref{table:power}.
Since we built our system for maximum output power and range, we did not make these hardware changes in this work, but we wish to outline the available design choices for use-cases demanding lower power consumption.}

\vspace{2mm}
\subsection{Cost Analysis}

The \shortname reader is designed with the goal of simplifying the deployment of backscatter technology to unleash the untapped potential of backscatter. Cost plays a critical role in achieving this objective. 
Table~\ref{table:cost} outlines the cost structure of the different components of the system and compares it with a legacy HD LoRa backscatter reader. {\revf Our analysis using pricing data from Octopart~\cite{octopart}} shows that at low volumes of 1,000 units, the FD reader costs \$27.54, only 10\% more than the cost of two HD readers. We believe that further optimization and leveraging economies of scale, coupled with the reuse of radios and processing power upon integration with existing devices such as IoT gateways, smartphones, and tablets, can further reduce the {\rev solution} cost.

\vspace{-2mm}
\subsection{LoRa Backscatter Tag}
The LoRa backscatter tag used in this work is based on the design proposed in~\cite{lorabackscatter}. The LoRa baseband and subcarrier chirp-spread-spectrum-modulated packets are generated using Direct Digital Synthesis (DDS) on an AGLN250 Igloo Nano FPGA~\cite{AGLN250}. The output of the FPGA is connected to SP4T ADG904 RF switch~\cite{adg904} to synthesize single-side-band backscatter packets. The backscatter tag design also incorporates an On-Off Keying (OOK) based wake-on radio with sensitivity down to -55~dBm and an ADG919~\cite{adg919} SPDT switch to multiplex a 0~dBi omnidirectional PIFA between the receiver and the backscatter switching network. The total loss in the RF path (SPDT + SP4T) for backscatter is $\sim$~5~dB.

\begin{table}[t]
\centering
\begin{threeparttable}
\centering
\caption{Estimated Power Consumption of \shortname Reader}
\begin{tabular}{|r | c | r |} 
 \hline
 \rowcolor{lightgray} TX Power & Applications & Peak Power \\ [0.5ex] 
 \hline 
 30~dBm & Plugged-in devices & 3,040~mW\tnote{+} \\
 \hline
 20~dBm & Laptops, Tablets & 675~mW \\
 \hline
 10~dBm & Phones, Battery Packs & 149~mW \\ 
 \hline
 4~dBm & Phones, Battery Packs & 112~mW \\ 
 \hline
\end{tabular}
\begin{tablenotes}\footnotesize
\item[+] Measured result.
\end{tablenotes}
\label{table:power}
\end{threeparttable}
\vskip -0.1 in
\end{table}


\begin{table}[tbp]
\vskip -0.1in
\centering
\caption{Cost Analysis of FD \& HD Backscatter Hardware}
\normalsize
\begin{tabular}{|c|c|c|} 
 \hline
 \rowcolor{lightgray} 
 Components & FD Cost & (2$\times$)~HD Cost\\ [0.5ex] 
 \hline 
  Transceiver & \$4.16 & (2$\times$)~\$4.16\\  
 \hline
  Synthesizer & \$7.15 & N/A \\ 
 \hline
   Power Amplifier & \$1.33 & (2$\times$)~\$1.33 \\ 
 \hline
  Cancellation Network & \$5.78 & N/A\\ 
 \hline  
  MCU & \$1.70 & (2$\times$)~\$1.30 \\ 
 \hline 
  Power Management & \$2.25 & (2$\times$)~\$1.95 \\
 \hline 
  Passives & \$2.52 & (2$\times$)~\$1.54 \\
 \hline 
  PCB fabrication~\cite{pcbminions} & \$1.07 & (2$\times$)~\$0.79\\ 
 \hline 
  Assembly~\cite{pcbminions} & \$1.58  & (2$\times$)~\$1.38\\
 \hline
 \hline 
 \bf{Total} & \bf{\$27.54} & \bf{\$24.90}\\
 \hline
\end{tabular}
\label{table:cost}
\vskip -0.2 in
\end{table}

\begin{figure*}[t!]
    \centering
    \begin{minipage}{0.23\linewidth}
        \centering
        \capsize
        \includegraphics[width=1\linewidth]
        {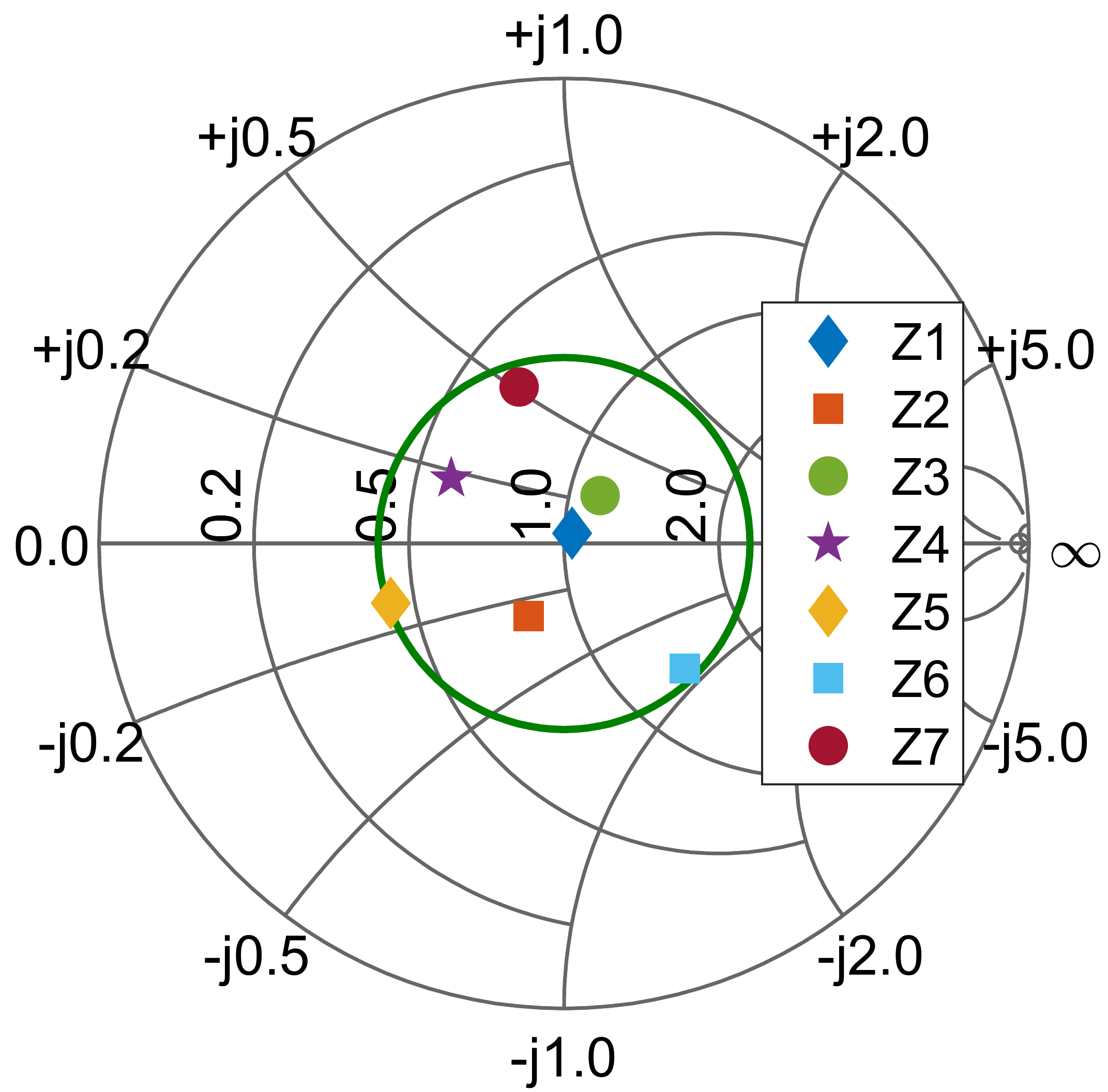}\\
         \vskip -0.05 in
        (a) Selected impedances.
        \vskip -0.05 in
    \end{minipage}
    \begin{minipage}{0.38\linewidth}
        \centering
        \capsize
        \includegraphics[width=1\linewidth]
        {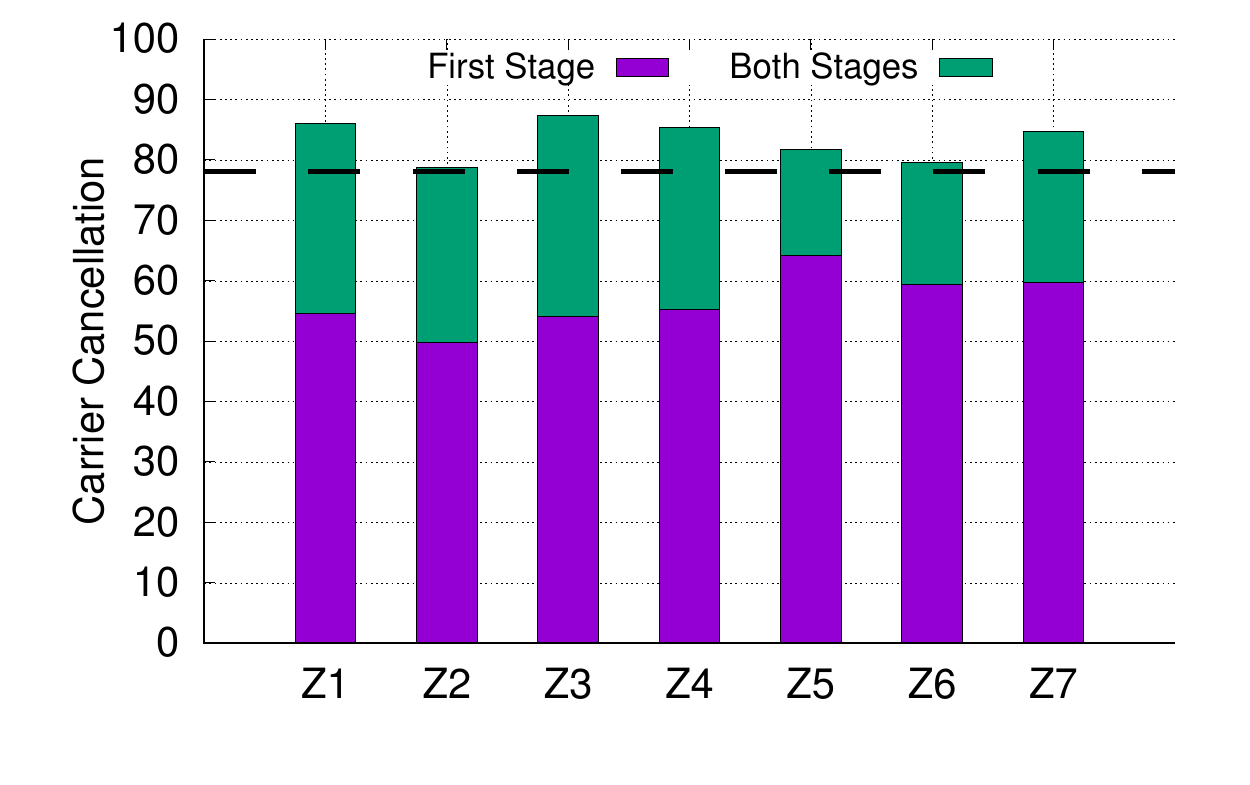}\\
        \vskip -0.2 in
        (b) First and second stage carrier cancellation.
        \vskip -0.05 in
    \end{minipage}
    \begin{minipage}{0.38\linewidth}
        \centering
        \capsize
        \includegraphics[width=1\linewidth]
        {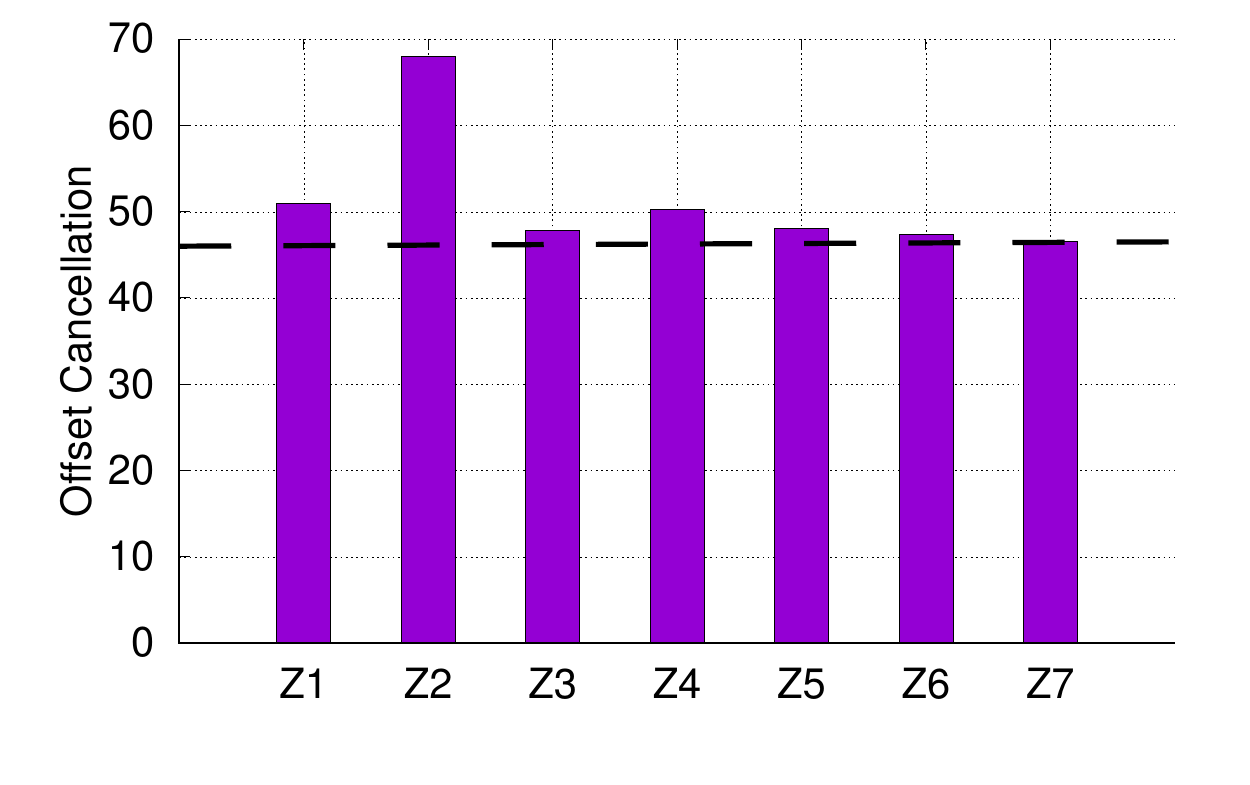}\\
        \vskip -0.2 in
        (c) Cancellation at 3MHz subcarrier offset.
        \vskip -0.05 in
    \end{minipage}
    \vskip -0.1 in
    \caption{SI cancellation as a function of variation in antenna impedance.}
    \label{fig:Can}
    \vskip -0.2 in
\end{figure*}
\section{Evaluation }
\label{sec:eval}

First, we validate our cancellation approach by measuring the carrier and offset cancellation of our novel two-stage impedance tuning network. 
Then, we measure the time overhead incurred by our tuning approach. 
Next, we evaluate the \shortname system performance in a wired setup to neutralize multi-path effects, followed by line-of-sight and non-line-of-sight wireless deployments. Finally, we measure the performance of the mobile version of our system. 

Unless mentioned otherwise, {\rev we set the subcarrier modulation frequency to 3~MHz,} and configure the tag to transmit 1,000 packets with $SF=12$, $BW=250kHz$, (8,4) Hamming Code with an 8-byte payload, a sequence number for calculating PER, and a 2-byte CRC. Additionally, we initiate uplink by sending a downlink OOK-modulated packet at 2~kbps to wake up the tag and align the tag's backscatter operation to the carrier. Downlink also enables channel arbitration between multiple tags, sending acknowledgments, packet re-transmissions, and other functionalities~\cite{lorabackscatter, passiveWiFi}. {\rev The} evaluation of these features is beyond the scope of this work.

\vspace{-2mm}
\subsection{Cancellation Network}

The cancellation network performance depends on the antenna impedance, which is sensitive to environmental variations (see~\xref{sec:approach}). To demonstrate that our system can achieve the required cancellation across a range of expected antenna impedances, we simulate antenna impedances with custom circuit boards with an 0402 footprint and an SMA connector. We populate the pads with discrete passives to represent antenna impedances with $0~\leq~|\Gamma|~\leq~0.4$. We test seven antenna impedances, as shown on the smith chart in Fig.~\ref{fig:Can}~(a).

To measure cancellation, we attach a board representing an antenna impedance to the antenna port of our \shortname reader with a Murata measurement probe~\cite{MM8430}. 
We disconnect the receiver and attach the receiver input to a spectrum analyzer using another RF probe. 
We set the transmitter to 915~MHz and 30~dBm output power. Since the receiver is disconnected, we {\rev cannot} \cmt{are unable to} measure RSSI and, {\rev therefore,} cannot utilize the tuning algorithm. 
We manually set the capacitor states in a two-step process similar to the tuning algorithm.
First, we fix the second-stage capacitors and manually tune the first stage for the best SI cancellation, then, we manually tune the second stage. 
Fig.~\ref{fig:Can}(b) shows the SI carrier cancellation results for one- and two-stage tunable impedance networks. Results show that a single stage is insufficient to achieve 78~dB carrier cancellation, whereas the two-stage design meets the specification. Next, we measure offset cancellation by keeping the same capacitor configuration and sweeping the carrier source between 905~-~925~MHz in 100~kHz frequency increments. Fig.\ref{fig:Can}(c) shows the offset cancellation for different antenna impedances at 3~MHz offset. Our results show that we achieve our target of 46.5~dB offset cancellation for all antenna impedances.

\vspace{-2mm}
\subsection{Tuning Overhead}

\begin{figure}[t!]
    \centering
    \includegraphics[width=1\linewidth]
    {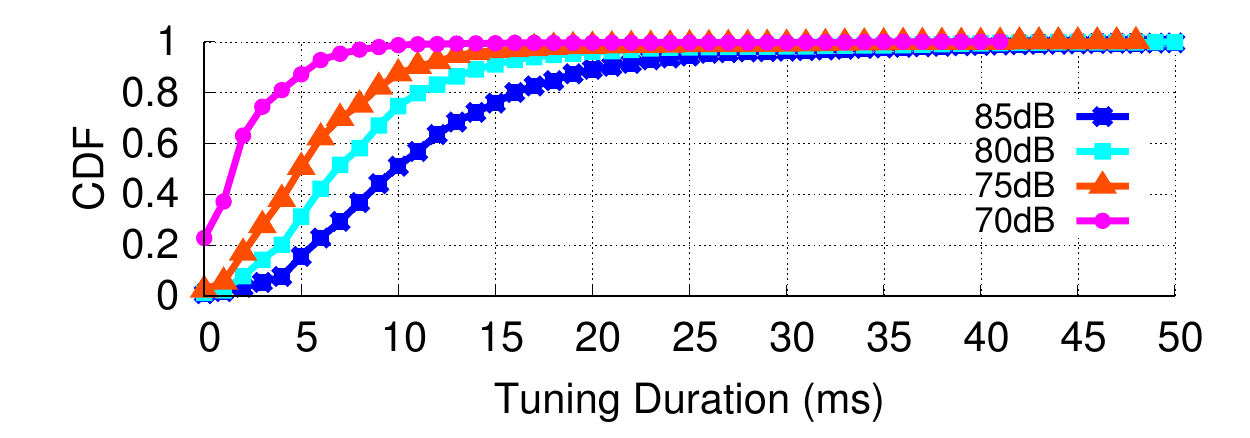}\\
    \vskip -0.2 in
    \caption{\textbf {Tuning algorithm overhead.} 
    We measure the overhead for different thresholds over 10000 packets.}
    \label{fig:tuning}
    \vskip -0.2 in
\end{figure}

To measure the performance of our tuning algorithm, we place the \shortname reader with the PIFA on a table in a typical office environment. 
We collect 10,000~packets from a tag placed 20~ft away over the course of 80~minutes with multiple people sitting nearby and walking in the vicinity of the system.
We modify the target SI cancellation threshold in the tuning algorithm to 70, 75, 80, and 85~dB and run experiments to measure the time required for tuning. 
The tuning algorithm was able to achieve the target SI in 99\% cases. 
We plot the CDF of tuning overhead for different cancellation thresholds in Fig.~\ref{fig:tuning}. 
As expected, the tuning duration increases with {\rev the} target threshold. 
For a threshold of 80~dB, the average tuning duration is 8.3~ms, corresponding to an overhead of 2.7\%. 
The RSSI measurements from the SX1276 chipset are noisy, and we take the mean over 8~readings for each tuning step, which is the key limitation. 
Each tuning step takes about 0.5~ms, dominated by the SPI transactions and settling time of the receiver. 
An RF power detector, which is beyond the scope of this work, can be used to provide faster feedback at the expense of increased cost.

\vspace{-2mm}
\subsection{Receiver Sensitivity Analysis}

\begin{figure}[t!]
\includegraphics[width=\columnwidth]
{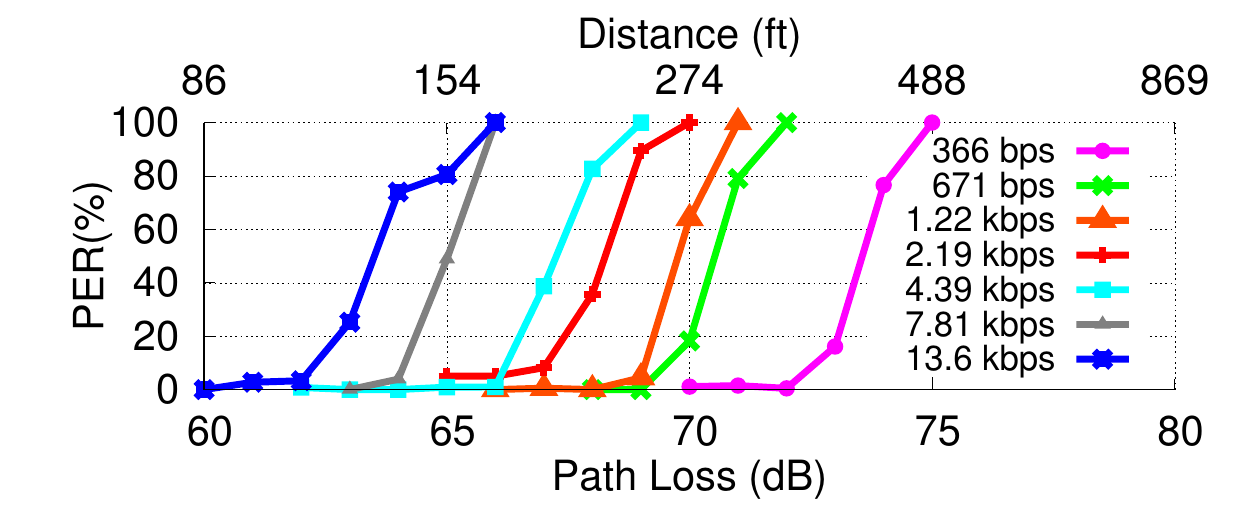}
\vskip -0.2 in
\caption{{\textbf{Receiver Sensitivity Analysis.} We plot the PER as a function of path loss for different data rates.}}
\label{fig:wired_test}
\vskip -0.15 in
\end{figure}

To evaluate the receive sensitivity of the \shortname system without the effect of multi-path signal propagation, we create an equivalent wired setup. We use RF cables and a variable attenuator to connect the antenna port of the \shortname reader to a LoRa backscatter tag. We vary the in-line attenuator to simulate path loss, corresponding to different operating distances between the reader and the tag. We start with an attenuator value at which we receive all packets and then slowly increase the attenuation until no packets are received. We configure the SF and BW parameters to cover a range of sensitivity and data rates. 

Fig.~\ref{fig:wired_test} plots PER as a function of path loss in a wired setup for different data rates. Since sensitivity is inversely proportional to data rate, lower data rates can operate at higher path loss, which translates to longer operating distances. For a $PER \leq 10\%$, the expected LOS range at the lowest data-rate of 366~bps {\rev ($SF = 12, BW = 250~KHz$)} is 340~ft, with the range decreasing successively for higher bit rates, down to 110~ft for 13.6~kbps {\rev ($SF = 7, BW = 500~KHz$)}. 


\begin{figure}[t!]
    \centering
    \begin{minipage}{1\linewidth}
        \centering
        \includegraphics[width=1\linewidth]
		{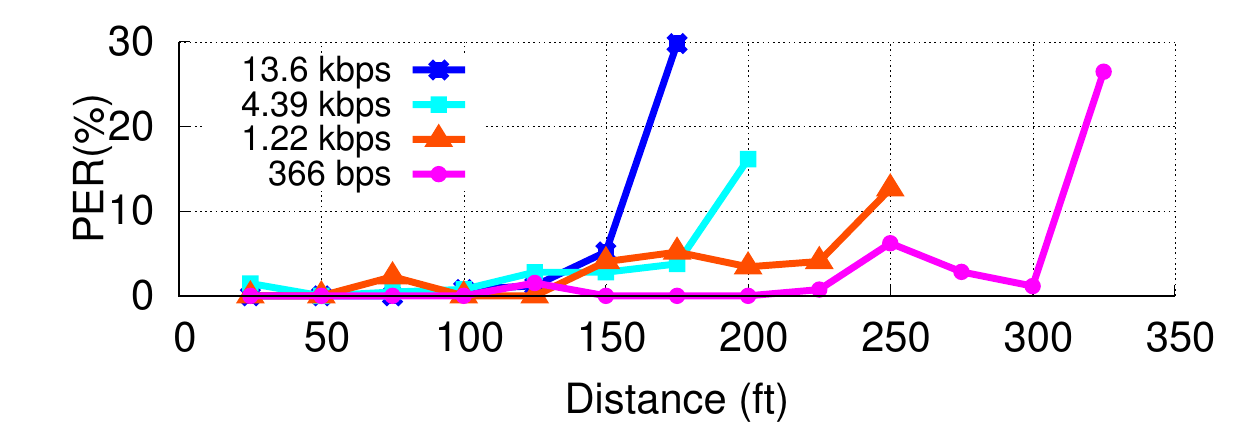}\\
		\vskip -0.075 in
        \capsize{(a) PER vs Distance for various data-rates.}
        \vskip -0.1 in
        \hfill
    \end{minipage}
    \vskip 0in
    \begin{minipage}{1\linewidth}
        \centering
        \includegraphics[width=1\linewidth]
        {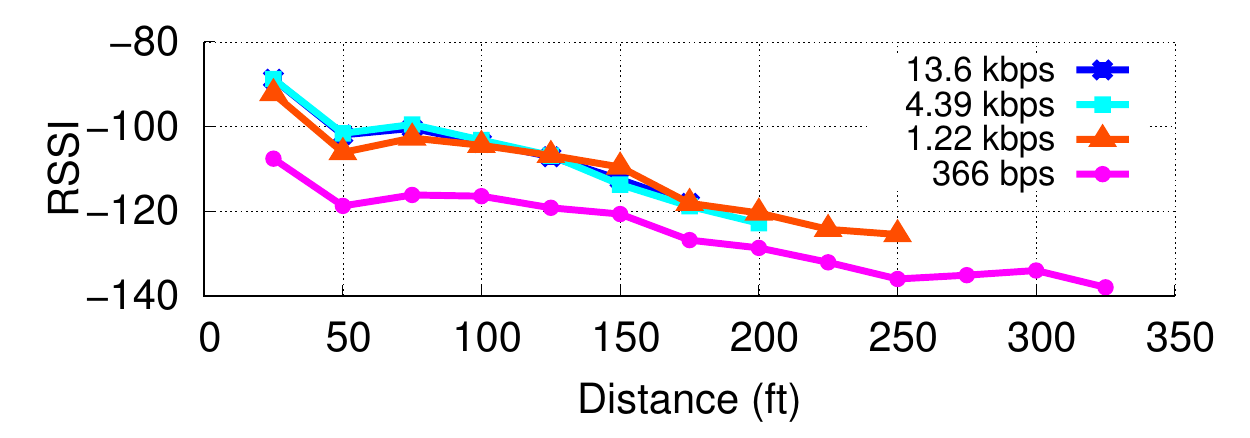}\\
        \vskip -0.075 in
        \capsize{(b) RSSI vs Distance for various data-rates.}
        \vskip -0.05 in
    \end{minipage}
    \vskip -0.1 in
    \caption{\textbf{Line-of-Sight Wireless Tests.} We move the backscatter tag away from the reader in line of sight.}
    \label{fig:LOS}
    \vskip -0.2 in
\end{figure}

\vspace{-2.5mm}
\subsection{Line-of-Sight (LOS) Wireless Testing}

We deploy the \shortname system \cmt{in an open field} in a nearby park to measure LOS performance. 
{\rev For best performance, we configure the reader as a base-station (see~\xref{sec:power}) by connecting an 8~dBiC circularly polarized patch antenna~\cite{patchantenna}, placed on a 5~ft tall stand, to the antenna port and set transmit power to 30~dBm.} 
We place the tag at the same height and move it away in 25~ft increments. Fig.~\ref{fig:LOS} plots PER and RSSI as a function of distance for four different data rates. 
Our results show that, for $PER < 10\%$, at the lowest data rate, the system can operate at a distance of up to 300~ft with a reported RSSI of -134~dBm. 
For the highest data rate, the operating distance was 150~ft {\rev at -112~dBm RSSI}. 

A prior HD LoRa backscatter system reported a range of 475~m between the two radios~\cite{lorabackscatter}; this corresponds to a range of 780~ft for an FD system. 
Our FD system achieves a shorter range and this can be attributed to two factors. 
First, the HD system evaluation uses a -143~dBm sensitivity protocol at 45~bps versus the -134~dBm sensitivity at 366~bps used in this work.
The 45~bps packets are 2.4~sec long, 6~$\times$ the FCC maximum channel dwell time (see~\xref{sec:system_lora_primer}).
Additionally, the FD system uses a hybrid coupler architecture. This reduces cost, but incurs a 7~dB loss (see~\xref{sec:impl}).
So, in total, our link budget is reduced by 16~dB. This translates to a 2.5~$\times$ range reduction, close to the 300~ft range of our system.


\begin{figure}[t!]
    \centering
    \begin{minipage}{1\linewidth}
        \centering
        \capsize
		\includegraphics[width=0.8\linewidth]
        {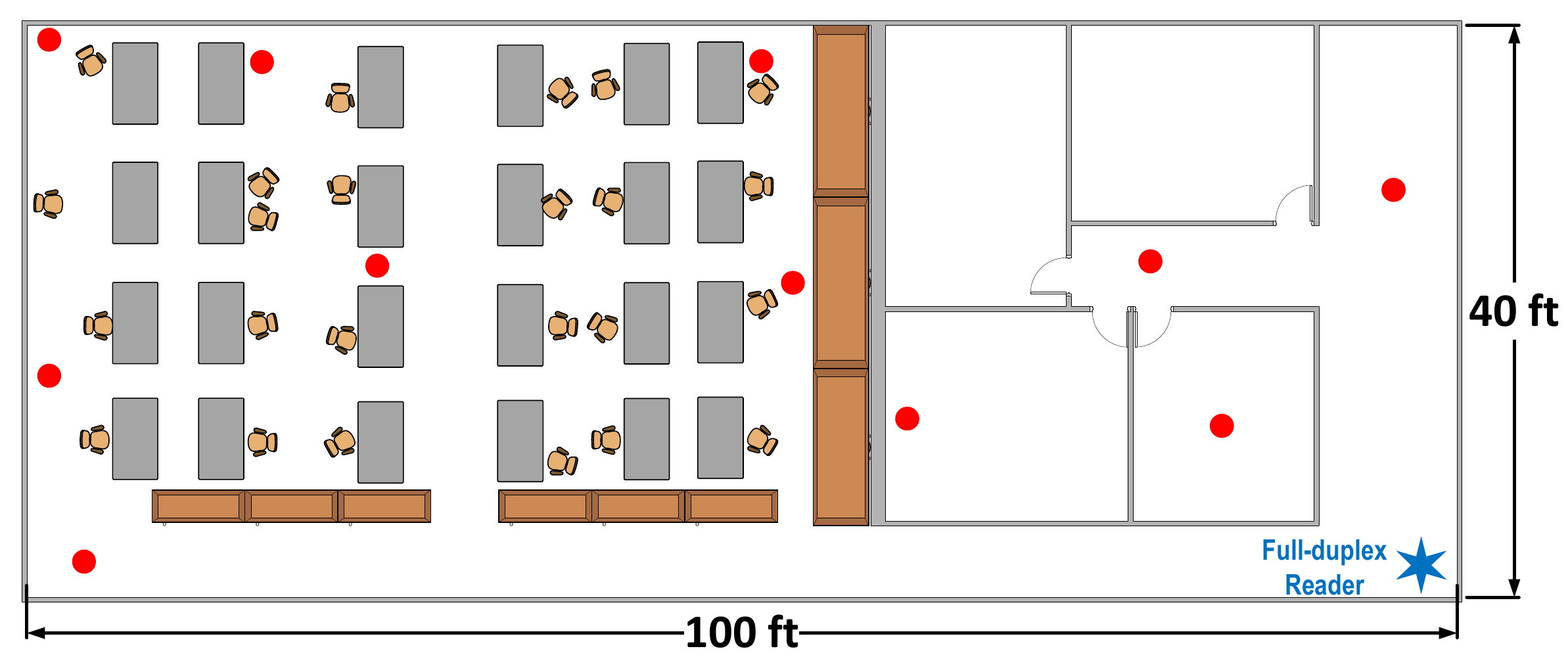}\\
        \vskip -0.05 in
        \capsize{(a) Floor plan of the 4,000~$ft^{2}$ office space.}
        \vskip -0.1 in
        \hfill
    \end{minipage}
    \begin{minipage}{1\linewidth}
        \centering
        \capsize
        \includegraphics[width=1\linewidth]
        {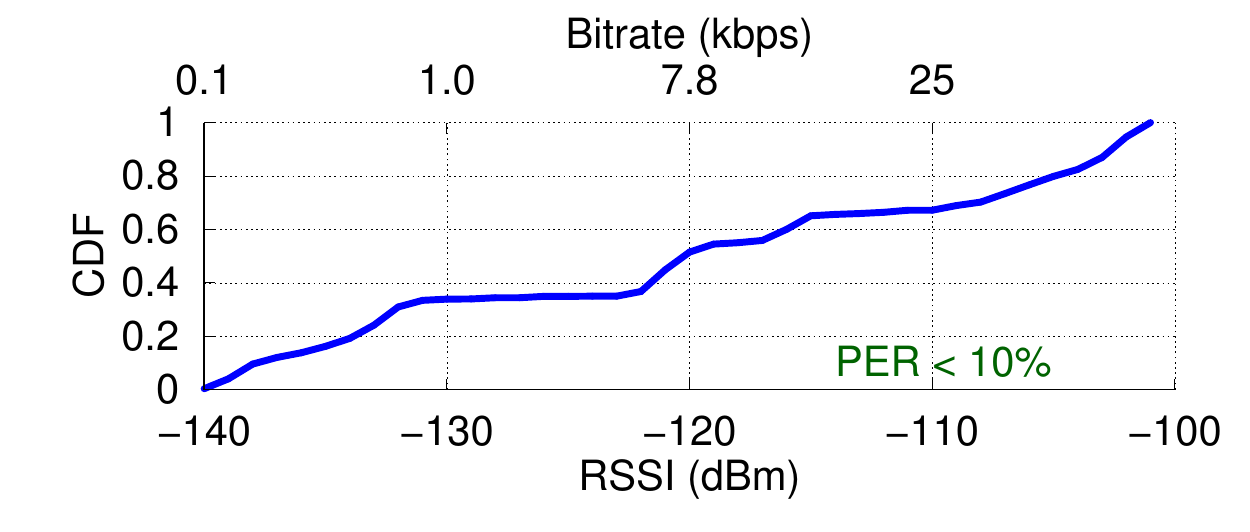}\\
        \vskip -0.075 in
        (b) RSSI of all measurements in the office space.
        \vskip -0.05 in
    \end{minipage}
    \vskip -0.1 in
    \caption{\textbf{Non-Line-of-Sight Wireless Tests.} We place the backscatter tag at 10 locations shown as red dots.}
    \vskip -0.25 in
\label{fig:NLOS}
\end{figure}

\vspace{-2mm}
\subsection{Non-Line-of-Sight (NLOS) Wireless Tests}

Next, we set up in a 100~ft~$\times$~40~ft office building to evaluate performance in a more realistic NLOS scenario.
We place the {\rev base-station} reader in {\rev one} corner of the building and move the tag to ten locations to measure performance through cubicles,\cmt{through} multiple concrete and glass walls, and down hallways.
The floor plan of the building is shown in Fig.\ref{fig:NLOS}(a). 
The blue star in the lower-right corner indicates the position of the FD reader, and the red dots indicate the different locations of the tag throughout the office space. 
We transmit 1,000~packets at each location, and a CDF of the aggregated RSSI data from the test is shown in Fig.\ref{fig:NLOS}(b). 
We observed a median RSSI of -120~dBm and PER of less than 10\% at all the locations demonstrating that the \shortname system is fully operational in the office space with a coverage area of 4,000~ft$^2$. 

\begin{figure*}[!t]
    \centering
     \vskip -0.05 in
    \begin{minipage}{0.24\linewidth}
        \centering
        \vskip 0.1 in
        \capsize
        \includegraphics[width=1\linewidth]
        {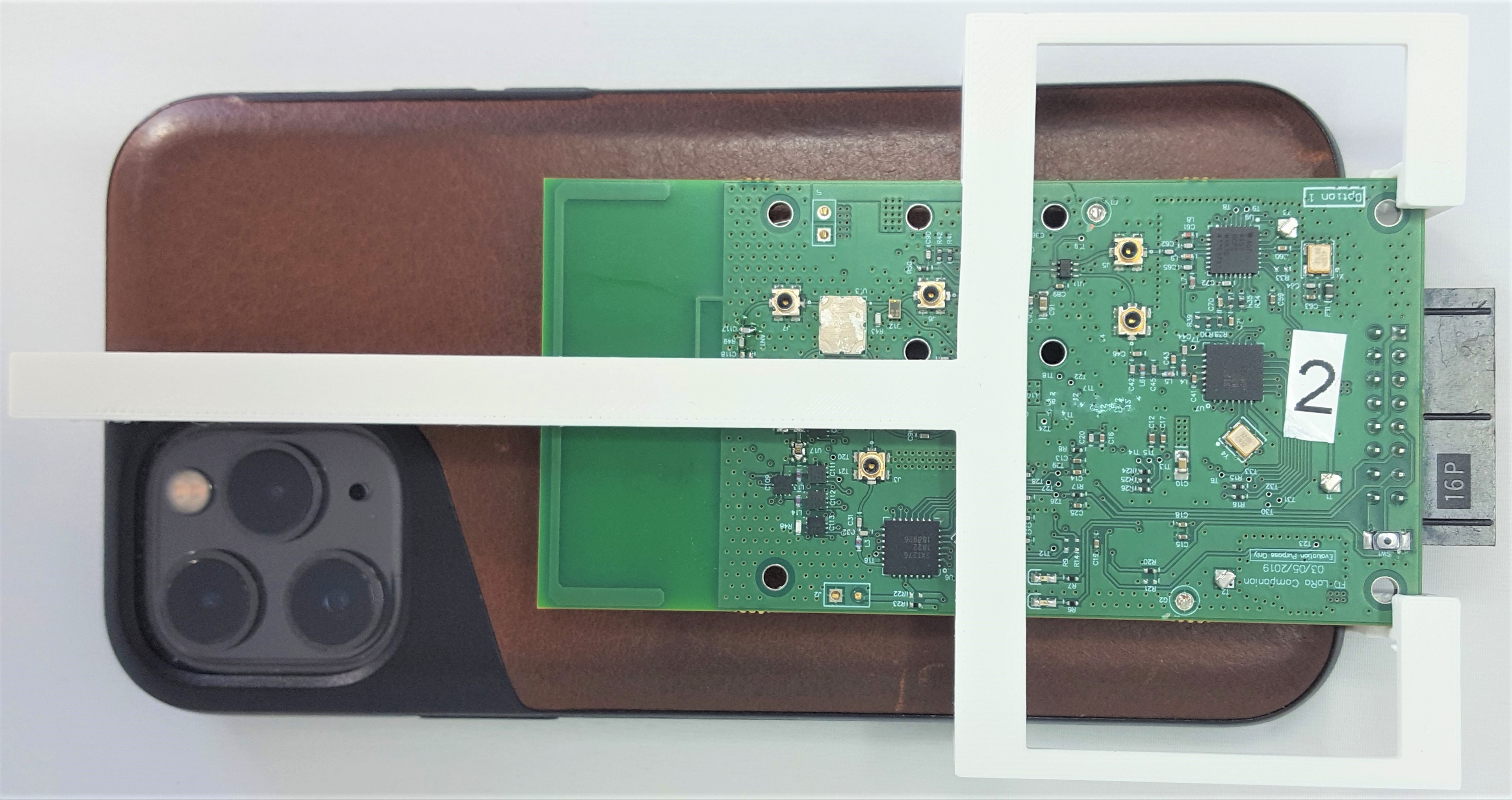}
        \vskip 0.1 in
        (a) \shortname on the back of a smartphone.
    \end{minipage}
    \begin{minipage}{0.36\linewidth}
        \centering
        \capsize
		\includegraphics[width=1\linewidth]
        {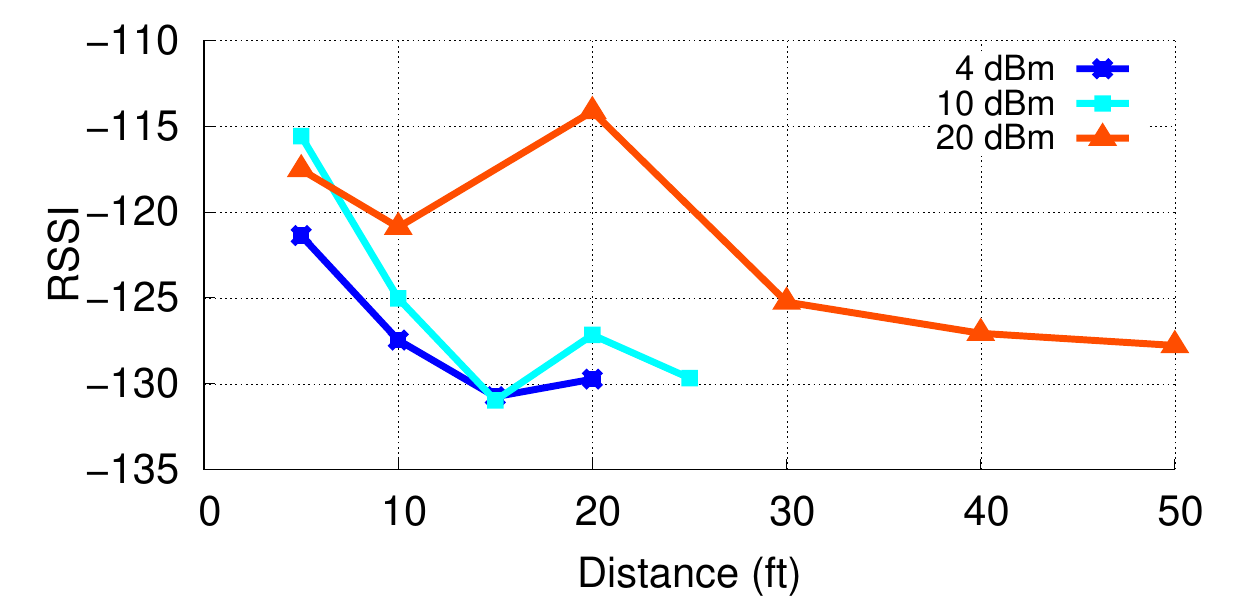}\\
        \vskip -0.075in
        (b) RSSI vs Distance for line-of-sight testing.
    \end{minipage}
    \begin{minipage}{0.36\linewidth}
        \centering
        \capsize
        \includegraphics[width=1\linewidth]
        {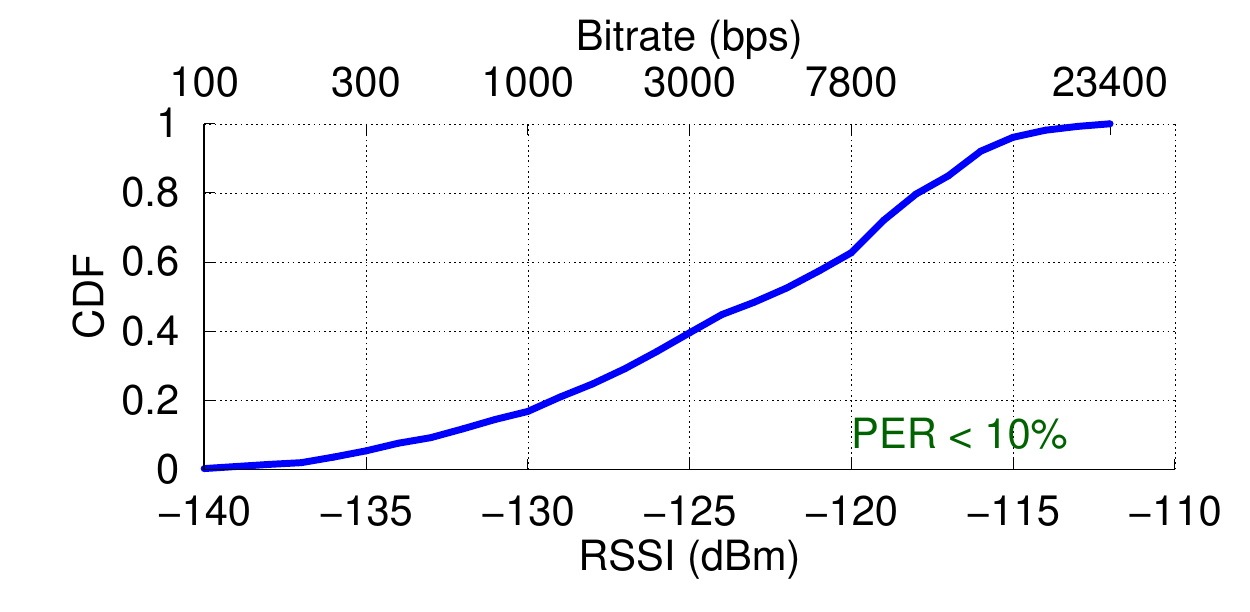}\\
        \vskip -0.075in
        (c) Packet RSSI with the mobile reader in a user's pocket moving around a table. 
    \end{minipage}
    \vskip -0.15 in
    \caption{Integration with Mobile Devices. 
}
    \label{fig:Phone}
    \vskip -0.2 in
\end{figure*}

\subsection{Integration with Mobile Devices}

\label{sec:mobile}
{\rev Finally, we evaluate the performance of the mobile version (see~\xref{sec:power}) of the FD reader.
We attach the mobile reader to the back of an iPhone 11 Pro, as shown in Fig.~\ref{fig:Phone}(a) and \cmt{We use the on-board antenna and place the phone face down on a wooden table.} configure the reader to transmit at 4~dBm, 10~dBm, and 20~dBm to resemble the transmit power of mobile devices. We move a backscatter tag away from the reader in 5~ft increments until PER~>~10\%.} Fig.~\ref{fig:Phone}(b) plots the RSSI of the received packets as a function of distance. The plots show that at 4~dBm, the mobile reader operates up to 20~ft and the range increases beyond 50~ft (the length of the room and limit of our testing) for a transmit power of 20~dBm. These distances are sufficient for connecting peripheral, wearable, and medical devices to a smartphone using backscatter at extremely low cost, small size, and low power consumption. {\rev These experiments were done in an uncontrolled wireless environment and the variation in signal strength at different locations is due to multi-path effects, which is typical of practical wireless testing.}

To demonstrate that our \cmt{tuning algorithm} {\rev system} can adapt to variations in environment and antenna impedance, we place the \shortname enabled smartphone in a subject's pocket and set the transmit power to 4~dBm. We place a tag at the center of an 11~ft~$\times$~6~ft table, and the subject walks around the perimeter of the table, receiving more than 1,000~packets. The performance is reliable with PER~<~10\%, which demonstrates the efficacy of our tuning algorithm. Fig.~\ref{fig:Phone}(c) plots the CDF of RSSI for all the packets. The backscatter tag measures 2~in~$\times$~1$\frac{1}{2}$~in, resembling the size of a pill bottle. This demonstrates that a mobile \cmt{\shortname system integrated into a} smartphone can use backscatter to communicate with a prescription pill bottle or insulin pen, allowing tracking of medication and drug delivery.

\section{Applications}
\label{sec:app}


\begin{figure*}[!t]
    \centering
     \begin{minipage}{0.24\linewidth}
        \centering
        \vskip 0.05 in
        \includegraphics[width=1\linewidth]
        {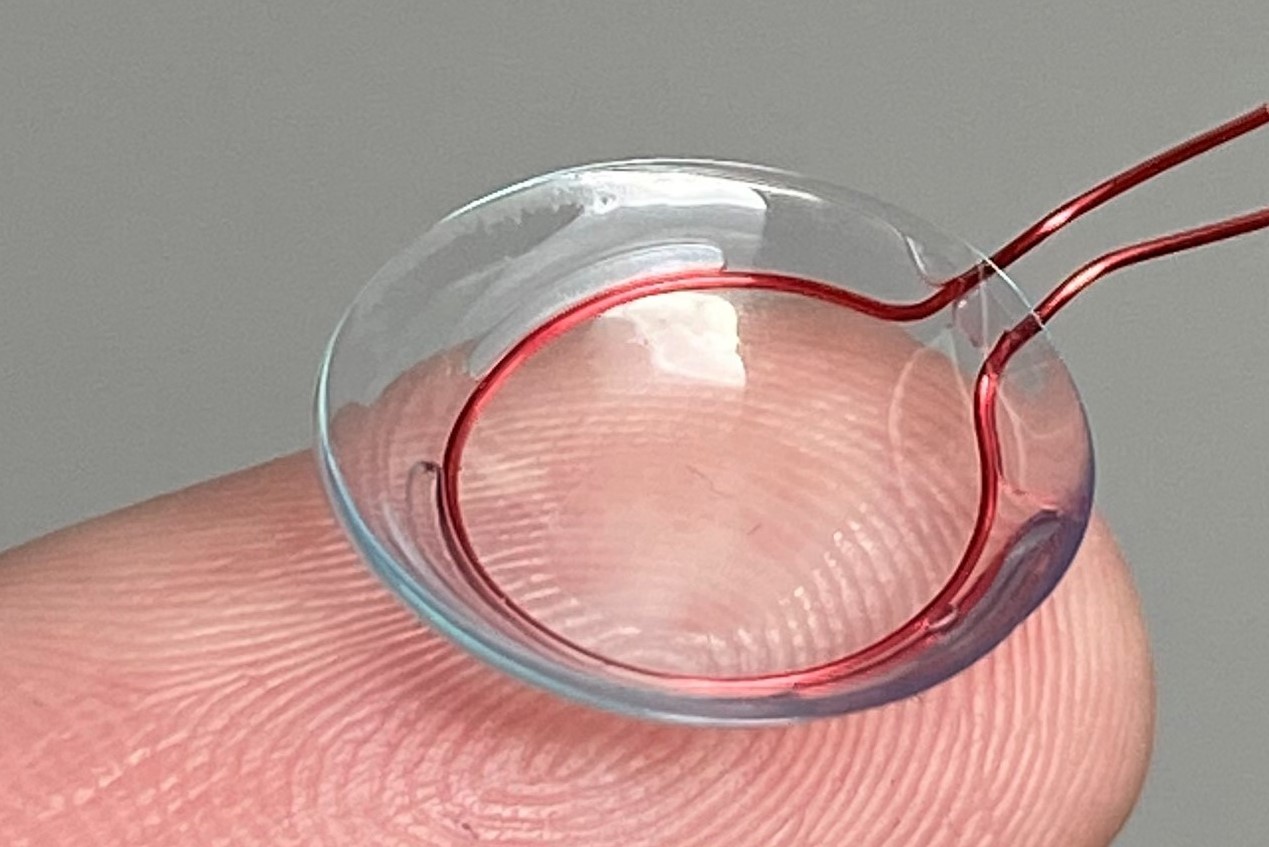}\\
        \vskip 0 in
        \capsize{(a) Contact lens antenna prototype.}
        \vskip -0.10 in
    \end{minipage}
    \begin{minipage}{0.36\linewidth}
        \centering
        
		\includegraphics[width=1\linewidth]
        {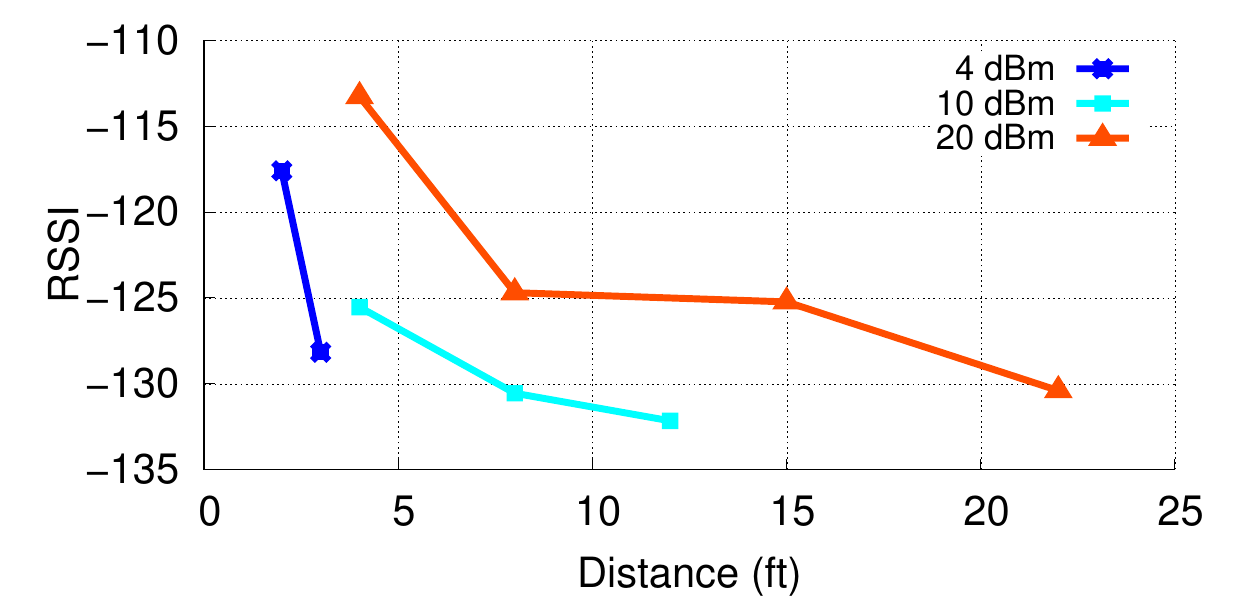}\\
        \vskip -0.05 in
        \capsize{(b) RSSI vs Distance for different transmit power.}
        \vskip -0.10 in
        \hfill
    \end{minipage}
    \begin{minipage}{0.36\linewidth}
        \centering
        
        \includegraphics[width=1\linewidth]
        {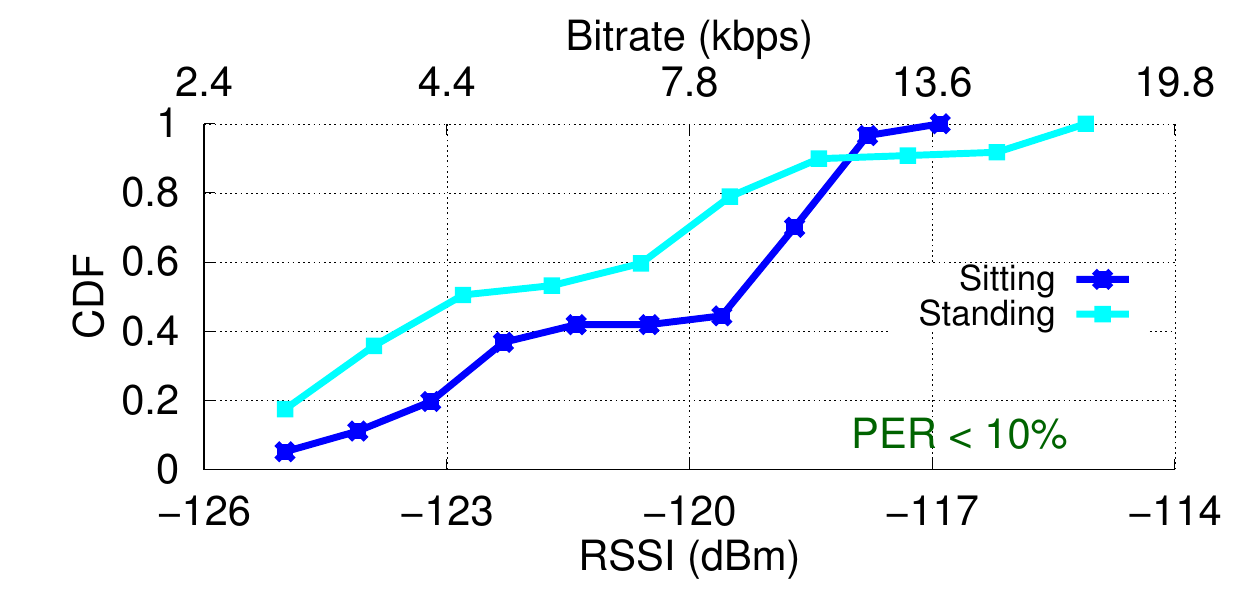}\\
        \vskip -0.05 in
        \capsize{(c) RSSI when reader is inside a user's pocket.}
        \vskip -0.10 in
    \end{minipage}
    \vskip -0.10 in
    \caption{A mobile \shortname reader communicating with a contact lens prototype.}
    \label{fig:ContactLens}
    \vskip -0.2 in
\end{figure*}

We demonstrate two applications for our FD system. First, we show how a mobile reader can collect data from a smart contact lens, a particularly challenging RF environment. Next, we demonstrate a precision agriculture application by mounting the reader to the bottom of a drone, which can be flown over farms and use backscatter to collect data from sensors distributed in a field. The use of a single reader coupled with a highly sensitive long-range backscatter protocol enables these applications, even in these challenging deployments.

\subsection {Contact Lens}
We use the mobile \shortname system mounted on the back of a smartphone to communicate with a backscatter tag equipped with a smart-contact-lens-form-factor antenna. {\rev We use the same backscatter endpoint as with other tests, but we} cut off the original PIFA and replace it with a small loop antenna of 1~cm diameter made with 30~AWG enameled wire. The antenna is encapsulated in two contact lenses and filled with contact lens solution to simulate the RF environment of an eye-ball, as shown in Fig.~\ref{fig:ContactLens}(a). Due to its small size and the ionic environment {\rev of the contact solution,} the loop antenna has an expected loss of 15~-~20~dB. 

We place the smartphone on a table and configure the mobile reader to transmit at 4, 10, and 20~dBm and move the contact lens backscatter prototype away from the smartphone. Fig.~\ref{fig:ContactLens}(b) shows the RSSI as a function of distance for various transmit powers. We show that the mobile reader at 10~dBm and 20~dBm transmit power can communicate with the contact lens at distances of 12~ft and 22~ft respectively for PER~<~10\%. Next, we put the mobile reader transmitting at 4~dBm in a 6~ft tall subject's pocket and hold the contact lens prototype near the subject's eye to simulate a realistic use case. Fig.~\ref{fig:ContactLens}(b) plots the CDF of the RSSI of decoded packets when the user was standing and sitting on a chair. The plot shows reliable performance with PER~<~10\% and a mean RSSI of -125~dBm. This demonstrates the feasibility of using backscatter to provide continuous connectivity between a user’s phone and a smart contact lens. This RF-challenged application was made possible even at such a low transmit power due to the high receive sensitivity of the system.   

\subsection {Drone with an FD Backscatter Reader}
\begin{figure}[!t]
    \centering
    \begin{minipage}{0.39\columnwidth}
        \centering
        \includegraphics[width=1\linewidth]
        {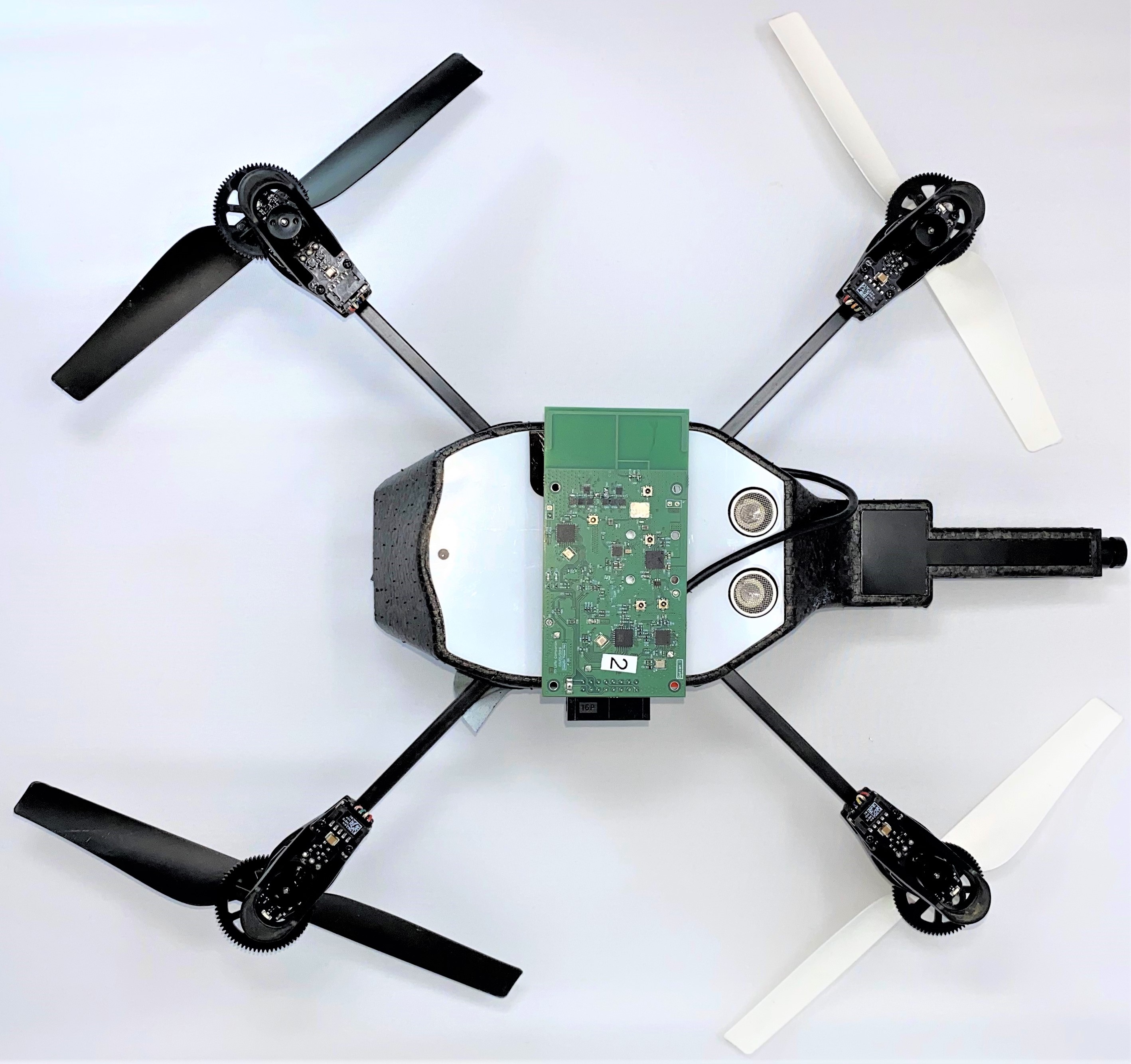}\\
        \capsize{(a) Reader mounted on the drone.}
        \vskip -0.05 in
    \end{minipage}
    \begin{minipage}{0.6\columnwidth}
        \centering
        \capsize
		\includegraphics[width=1\linewidth]
        {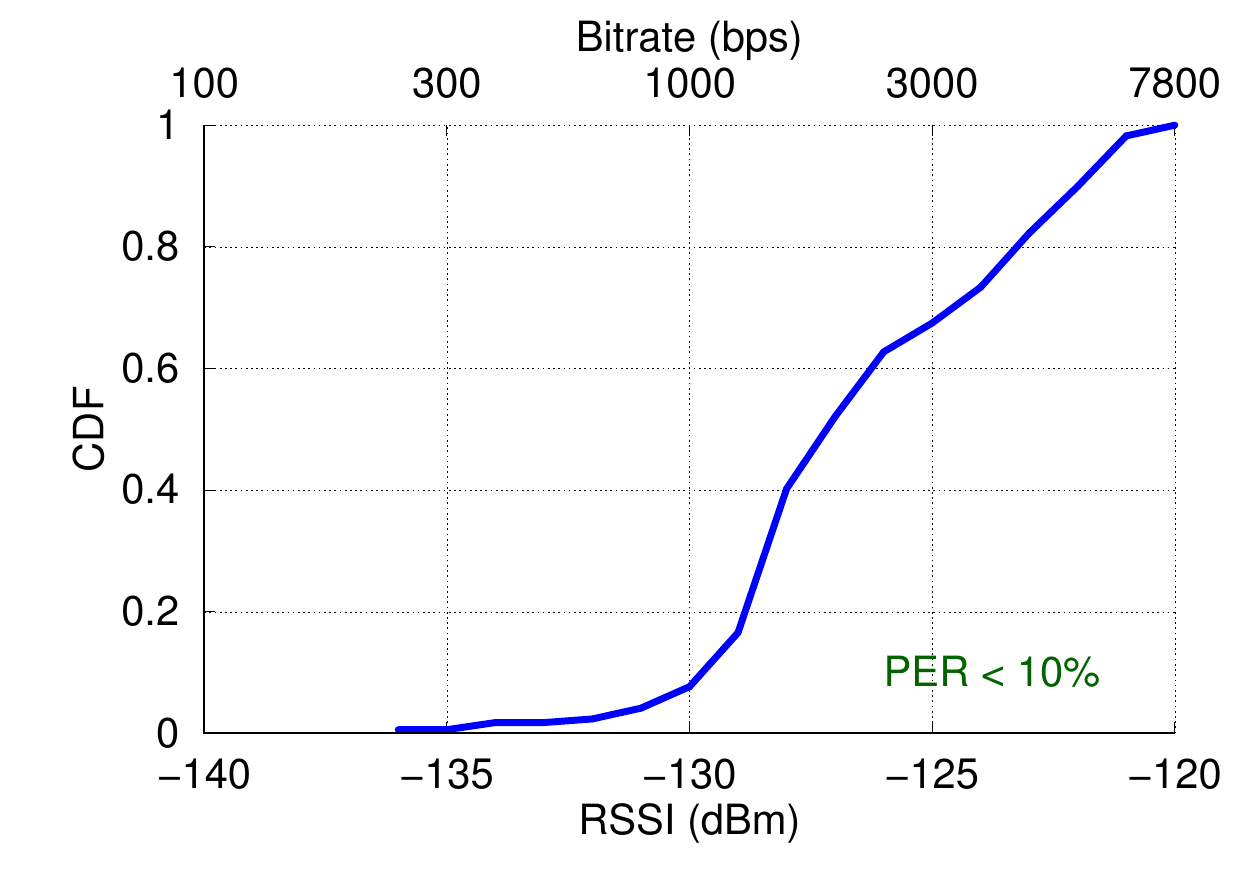}\\
        \capsize{(b) RSSI when drone is at 60~ft altitude.} 
        \vskip -0.05 in
        \hfill
    \end{minipage}
    
    \vskip -0.15 in
    \caption{ Backscatter enabled Low-Cost Drone.}
    \label{fig:Drone}
    \vskip -0.15 in
\end{figure}

Drones are extensively used for aerial surveillance in precision agriculture~\cite{Farmbeats}. We demonstrate how one can augment a drone's functionality by adding a \shortname reader to communicate with sensors distributed in a field using backscatter. We attach the mobile version of our reader to the bottom of a low-cost, commercially-available Parrot AR.Drone 2.0 quadcopter (<\$80)~\cite{ARdrone}, as shown in Fig.\ref{fig:Drone}(a). We power the reader from the drone's battery using a USB connector to demonstrate the ease of {\rev integrating} \cmt{integration of} our system. We set the transmit power to 20~dBm to reduce the burden on 7.5~Whr battery of the cheap drone. 
We place the tag on the ground simulating an agriculture sensor and fly the drone at a height of 60~ft. {\rev Due to practical challenges in accurately positioning the drone, we allow the drone to fly laterally during the test up to 50~ft from the center, which results in 80~ft maximum separation from the tag. This corresponds to an instantaneous coverage area of 7,850~ft$^2$. We collect more than 400 packets over 4 minutes with the drone flying around the coverage zone while keeping its altitude constant.}

Fig.\ref{fig:Drone}(b) plots CDF of the RSSI of the packets received by the drone {\rev over the entire duration of the test} for a PER <10\%. With a minimum of -136~dBm and median of -128~dBm, this demonstrates good performance for the area tested. With a flight time of 15~min and a top speed of 11~m/s, our cheap drone could, in theory, cover an area greater than 60~acres on a single charge. With a more powerful drone with higher payload capacity and longer flight time, one could integrate a higher gain antenna and transmit at higher power. This would result in a greater instantaneous coverage area and, with longer flight time, could achieve many times greater coverage on a single charge.

\section{Related Work}
\label{sec:related}

Our work is related to prior efforts in HD backscatter, FD backscatter, and in-band FD communication.

\vskip 0.05in\noindent{\bf Half-Duplex Backscatter.} Our work builds on recent efforts in developing backscatter solutions that are compatible with existing wireless standards such as Bluetooth~\cite{Ensworth2017,InterTechnology, freerider, relacks}, WiFi~\cite{passiveWiFi,InterTechnology, witag,spatialstream, freerider, on-body}, Zigbee~\cite{passive-zigbee}, and LoRa~\cite{lorabackscatter, plora} using half-duplex architectures. The backscatter endpoint is based on prior LoRa backscatter design~\cite{lorabackscatter}, but we take the next step of integrating the single-tone carrier source and LoRa receiver into a single device. \cmt{Although our design is built for LoRa protocol and receivers, the proposed full-duplex architecture and techniques are also applicable to WiFi, Bluetooth, Zigbee, and other protocols {\rev if subcarrier modulation is used to generate packets from a single-tone carrier.}}

In addition to standards-compliant backscatter, proprietary-protocol communications~\cite{Netscatter,OFDMAbackscatter,25MbpsBrain,polymorphic} (to improve data rates and throughput), applications such as wireless video streaming~\cite{batteryfreevideo,WiFiCam}, indoor localization~\cite{Sub-Centimeter,real-timeLoc,3DBackLoc}, and human activity recognition~\cite{BARNET,BTTN,WispActivity} have been realized with HD deployments. The techniques presented in this work can be extended to build an FD version of these systems.

\begin{table*}[t]
\begin{threeparttable}
\centering
\normalsize
\caption{Comparison of State of the Art Analog SI Cancellation Techniques}
\begin{tabular}{|c | c | c | c | c | c | c | c | c |}
 \hline
 \rowcolor{lightgray} 
  Reference & Cancellation Technique & \makecell{TX\\~~Signal~~} & \makecell{RX\\~~~~~Signal~~~~~} & \makecell{Analog \\Can.} & \makecell{TX\\~~Power~~} &  \makecell{Active \\Comp.\tnote{+}} & Size & Cost\\ [0.5ex] 
 \hline 
 ~\cite{duarte_design_2014} & \makecell{Multiple antenna + \\ auxiliary can. path} & \makecell{WiFi \\Packet} & \makecell{WiFi \\Packet}& 65~dB & 8~dBm & Yes &  \makecell{37~cm Ant. \\Separation} & High\\
 \hline 
 ~\cite{Chen2019} & \makecell{Circulator + 2 tap \\freq. domain equalization} & \makecell{WiFi \\Packet} & \makecell{WiFi \\Packet}& 52~dB & 10~dBm & Yes & $1.5~\times~4.0~$cm$^2$ & High\\
 \hline 
 ~\cite{korpi_full-duplex_2016}& \makecell{Circulator + 3 complex-\\ tap analog FIR filter} & \makecell{WiFi \\Packet} & \makecell{WiFi \\Packet} & 68~dB & 8~dBm & Yes & N.A. & High\\
 \hline 
~\cite{Chu2018} & \makecell{EBD + Double \\RF adaptive filter} & General & General & 72~dB &  12~dBm & Yes & \multicolumn{2}{c|}{Custom ASIC\tnote{*}} \\
\hline 
~\cite{Reiskarimian2018} & \makecell{Magnetic-free N-path \\filter-based Circulator} & General & General & 40~dB & 8~dBm & No & \multicolumn{2}{c|}{Custom ASIC\tnote{*}}\\
\hline 
~\cite{liempd_70_2016} & \makecell{EBD + passive \\tuning network} & General & General & 75~dB & 27~dBm & No & \multicolumn{2}{c|}{Custom ASIC\tnote{*}}\\
\hline
 ~\cite{backfi2015} & \makecell{Circulator + \\ 16 tap analog FIR filter} & \makecell{WiFi \\Packet} & \makecell{WiFi \\Backscatter} & 60~dB & 20~dBm & No & $10~\times~10~$cm$^2$ & High\\
 \hline 
~\cite{BLE-Monostatic} & \makecell{20dB Coupler + \\Active tuning network} & CW & \makecell{BLE \\Backscatter} & 50~dB & 33~dBm & Yes & N.A. & High\\
\hline 
~\cite{sdr_rfid}& \makecell{10dB Coupler + Atten.\\ + passive tuning network} & CW & \makecell{EPC \\ Gen 2} & 60~dB & 26~dBm & No &  $2.7~\times~2.0~$cm$^2$ & Low\\
 \hline 
 This Work & \makecell{Hybrid Coupler + \\passive tuning network} & CW & \makecell{LoRa \\ Backscatter} & 78~dB &  30~dBm & No & $2.5~\times~0.8$~cm$^2$ & Low\\
 \hline 
\end{tabular}

\begin{tablenotes}\footnotesize
\item[+] Active components include phase shifters, vector modulators, amplifiers and transconductance amplifiers which can contribute additional noise.
\item[*] Custom ASICs are incompatible with COTS transceivers and are only viable and cost-efficient at scale.
\end{tablenotes}

\label{table:comp_fd}

\end{threeparttable}

\vskip -0.15 in
\end{table*}

\vskip 0.05in\noindent{\bf Full-Duplex Backscatter.} 
A BLE monostatic/FD backscatter system was introduced in~\cite{BLE-Monostatic} that uses a 20~dB coupler, phase shifter, and variable attenuator for SI cancellation. However, due to additional losses in the coupler and limited cancellation depth, the communication range  was limited to 3~m, even with 33~dBm of output power. In~\cite{backfi2015}, an FD backscatter system using an SDR platform with analog and digital cancellation was introduced where WiFi packets are used as the carrier and the tag backscatters proprietary BPSK, QPSK, and 16-PSK modulated packets which were decoded by the SDR up to a distance of 7~m. {\rev Due to the wide-band nature of WiFi carrier signals, ~\cite{backfi2015} needs wide-band SI cancellation. A circulator and a 10~cm~$\times$~10~cm custom PCB \cmt{board} with 16 variable-gain delay lines are added to the SDR platform for wide-band analog cancellation, increasing both the solution cost and size.} In contrast, the \shortname system uses commodity LoRa radios and passive components for cancellation and can receive standard LoRa packets up to a distance of 300~ft.

RFID readers are also full-duplex devices that transmit a single-tone carrier and receive backscattered packets from RFID tags. However, EPC Gen2 readers are bulky, expensive~\cite{st25ru3993, Speedway}, and cannot compete with economies of scale of standard protocols. {\rev The operating range of passive RFID readers is determined by the power-harvesting sensitivity, not by the backscatter communication link. RFID readers operate at relatively short distances, even if the tag is powered from an alternative energy source~\cite{pv-rfid}, due to poor receive sensitivity ($-90$~dBm)~\cite{Speedway,st25ru3993}}. In contrast, our \shortname system achieves much longer operating distances at significantly lower cost by leveraging a highly sensitive commodity LoRa receiver, cheap passive components, and a microcontroller for deep SI cancellation. Low-cost RFID readers based on directional couplers have also been investigated~\cite{Jung2011,Keehr2018}, but they suffer from high Rx insertion loss and lower SI cancellation depth, which reduces range. In~\cite{drone_relay}, a full-duplex drone relay is presented to extend the range of a fixed RFID reader. The topology requires an additional relay on a drone to extend the fixed RFID reader range to 50~m. In contrast, our FD reader can be mounted on a flying drone to cover a significantly larger area.


\vskip 0.05in\noindent{\bf In-Band Full-Duplex Radios.} In-band full-duplex (IBFD) radios simultaneously transmit and receive at the same frequency. Recent works have used a combination of isolation, analog cancellation, and digital cancellation techniques to suppress SI below the receiver noise floor~\cite{bharadia_full_2013,korpi_full-duplex_2016,duarte_design_2014,Chen2019}. 

Existing isolation techniques use two or more physically-separated antennas~\cite{everett_passive_2014,duarte_design_2014,Dinc2015,Kumar2019}, discrete circulators~\cite{bharadia_full_2013,korpi_full-duplex_2016,zhang_wideband_2018,Khaledian2018,Chen2019}, integrated circulators~\cite{Reiskarimian2018,Dastjerdi2019}, or hybrid junctions~\cite{liempd_70_2016, Chu2018} to isolate transmitter and receiver. {\rev The} use of multiple antennas increases form factor while achieving limited isolation. Discrete circulators~\cite{SKYFR-001400} are bulky and expensive. While recent advances in integrated circulators~\cite{Reiskarimian2018,Dastjerdi2019} are promising, these devices are unable to handle the 30~dBm output-power requirement. Hybrid junctions, realized using couplers~\cite{Jung2011,Keehr2018} (such as the 3~dB coupler used in this work) or electrical balance duplexers (EBD)~\cite{liempd_70_2016, Chu2018}, incur a loss, but result in small-form-factor and inexpensive solutions. However, existing {\rev solutions with COTS components} cannot achieve 78~dB of SI cancellation~\cite{Jung2011,BLE-Monostatic,Keehr2018}, whereas our proposed two-stage impedance tuning network can be used to {\rev achieve this deep cancellation required for building} a cheap, low-complexity, long-range FD reader. {\rev Furthermore, ~\cite{liempd_70_2016, Chu2018} show a path towards further reducing the cost and size at high volumes by integrating the hybrid junction in silicon.}

Analog feed-forward cancellers can be combined with isolation techniques to enhance SI cancellation depth {\rev and bandwidth}. Various feed-forward PCB and ASIC implementations based on vector modulation~\cite{broek_2015}, finite impulse response filters~\cite{bharadia_full_2013,korpi_full-duplex_2016,zhang_wideband_2018,Chu2018}, frequency domain equalization based RF filters~\cite{Chen2019}, N-path filters~\cite{zhou_integrated_2015}, and {\rev Hilbert transform equalization~\cite{sayed_full-duplex_2017}} have been proposed. However, these techniques require additional active circuitry, which has a limited power-handling capability and generates noise that increases the receiver noise floor~\cite{chu_integrated_2018}. Furthermore, COTS phase shifters~\cite{JSPHS-1000} and vector modulators~\cite{hmc630} are bulky and expensive, which substantially increases cost, complexity, and form factor. {\rev In this work, we utilize the two-stage tunable impedance network architecture to achieve the required 78~dB cancellation depth. Since the transmitter carrier signal is only single-tone, we do not need the feed-forward paths to improve the bandwidth. 

In Table~\ref{table:comp_fd}, we summarize the state-of-the-art techniques used for analog SI cancellation and compare them with our approach in terms of cancellation depth, transmit power-handling capability, size, and cost.} 

Finally, digital  cancellers are often used to further suppress the SI below the receiver noise floor~\cite{bharadia_full_2013,korpi_compact_2017,kiayani_adaptive_2018, Chen2019} {\rev using both linear and non-linear cancellation techniques~\cite{Katanbaf2019}.} Digital cancellation requires access to baseband IQ samples. This function is not available on commodity radio chipsets and is typically implemented on SDRs~\cite{bharadia_full_2013,sim_nonlinear_2017,Chen2019,Akeela2018}, FPGAs~\cite{Tsoeunyane2017}, or DSPs~\cite{Atomix}, which are prohibitively expensive. Instead, since our interference is a single-tone at a frequency offset from the receive channel, we leverage the inherent baseband filtering capabilities of the commodity receiver to suppress the SI at the offset frequency.

\vspace{-2mm}
\section{Conclusion}
\label{sec:conclusion}

We present the first low-cost, {\rev long-range}, small-form-factor \name design. We use a single-antenna, hybrid-coupler-based architecture and introduce a novel two-stage tunable impedance network to meet the stringent SI-cancellation requirements using only passive components and a microcontroller. We build hardware using COTS components and prototype proof-of-concept applications for a smart contact lens and backscatter enabled drone.

\vspace{-2mm}
\section*{Acknowledgments}
{\rev
The authors would like to thank Aaron Parks, Jonathan Hamberg, and Ye Wang for their help. We would also like to thank the anonymous reviewers as well as our shepherd, Prof. Lin Zhong for their helpful feedback on the paper.
}
\clearpage
\pagestyle{plain}
\bibliographystyle{plain}
\bibliography{backscatter,FD}

\end{document}